\documentclass[preprint,prd,showpacs,showkeys]{revtex4-1}

\usepackage{amsmath}
\usepackage{amssymb}
\usepackage{amsfonts}
\usepackage{graphicx}
\usepackage{axodraw4j}
\usepackage{appendix}

\begin{document}

\title{Maintaining Gauge Symmetry in Renormalizing Chiral Gauge Theories}
\author{Er-Cheng Tsai}
\email{ectsai@ntu.edu.tw}
\affiliation{Physics Department, National Taiwan University, Taipei, Taiwan}

\begin{abstract}
It is known that the $\gamma_{5}$ scheme of Breitenlohner and Maison (BM) in
dimensional regularization requires finite counter-term renormalization to
restore gauge symmetry and implementing this finite renormalization in
practical calculation is a daunting task even at 1-loop order. In this paper,
we show that there is a simple and straightforward method to obtain these
finite counter terms by using the rightmost $\gamma_{5}$ scheme in which we
move all the $\gamma_{5}$ matrices to the rightmost position before
analytically continuing the dimension. For any 1-loop Feynman diagram, the
difference between the amplitude regularized in the rightmost $\gamma_{5}$
scheme and the amplitude regularized in the BM scheme can be easily
calculated. The differences for all 1-loop diagrams in the chiral
Abelian-Higgs gauge theory and in the chiral non-Abelian gauge theory are
shown to be the same as the amplitudes due to the finite counter terms that
are required to restore gauge symmetry.
\end{abstract}

\maketitle

\affiliation{Physics Department, National Taiwan University, Taipei, Taiwan}

\section{Introduction}

It is generally accepted that the original $\gamma_{5}$ dimensional
regularization scheme proposed by 't Hooft and Veltman \cite{HV} and later
systematized by Breitenlohner and Maison \cite{BM} can be used to regulate and
renormalize chiral gauge theories in a rigorous manner. In this BM scheme,
$\gamma_{5}$ is maintained as
\begin{equation}
\gamma_{5}=i\gamma^{0}\gamma^{1}\gamma^{2}\gamma^{3} \label{g5def}%
\end{equation}
even when the space-time dimension $n$ departs from $4$. Such $\gamma_{5}$
anti-commutes with $\gamma^{\mu}$ for $\mu$\ in the first 4 dimensions but
commutes with $\gamma^{\mu}$ when the index $\mu$ falls beyond the first 4
dimensions. As a consequence, an identity such as $\gamma_{5}\gamma^{\mu
}=-\gamma^{\mu}\gamma_{5}$ in $n=4$ dimensional space no longer holds under
dimensional regularization when the polarization $\mu$ is continued beyond the
first 4 dimensions. The continuation to $n\neq4$ for the Lagrangian of a
theory with a gauge invariant 4 dimensional Lagrangian therefore depends on
how we express and continue the terms involving product of $\gamma_{5}$ and
$\gamma^{\mu}$ matrices in the Lagrangian. Furthermore, terms that are not
gauge invariant in the $n$ dimensional Lagrangian must vanish when
$n\rightarrow4$ and thus contain a factor of $\left(  n-4\right)  $ or a
$\gamma^{\mu}$ matrix with $\mu$ in the extra-4 dimensions. Such gauge variant
evanescent terms will contribute to the violation of gauge symmetry in the
perturbative calculation of the theory under dimensional regularization.

The breakdown of gauge symmetry in the BM scheme can be remedied by
introducing gauge variant local counter terms to restore the renormalized Ward
identities \cite{WTI} or BRST \cite{BRST} gauge symmetry
\cite{GB2,FER,FG,MS,SANR}. This procedure of removing spurious anomalies is
usually a complicated and tedious task even at the 1-loop order. C. P. Martin
and D. Sanchez-Ruiz \cite{MS}\ managed to successfully calculated the 1-loop
finite counter terms needed for restoring gauge symmetry of the chiral
non-Abelian gauge theory. For the chiral Abelian-Higgs theory, the 1-loop
finite counter terms were later obtained by D. Sanchez-Ruiz \cite{SANR} in
which the laborious calculations were handled by computer routines.

In this paper, we shall present a simple and straightforward method for
obtaining these finite counter terms. This is done with the help of the
rightmost $\gamma_{5}$ scheme \cite{ECSM} in which the dimension $n$ is
analytically continued after all the $\gamma_{5}$ matrices have been moved to
the rightmost position.

\section{The rightmost $\gamma_{5}$ Scheme}

For the QED theory, the identity%
\begin{equation}
\frac{1}{\not \ell +\not k  -m}\not k  \frac{1}{\not \ell -m}=\frac{1}{\ell
-m}-\frac{1}{\not \ell +\not k  -m} \label{qedwti0}%
\end{equation}
is the foundation that a Ward identity is built upon. For a gauge theory
involving $\gamma_{5}$, there is a basic identity similar to $\left(
\ref{qedwti0}\right)  $ for verifying Ward identities:%
\begin{equation}
\frac{1}{\not \ell +\not k  -m}\left(  \not k  -2m\right)  \gamma_{5}\frac
{1}{\not \ell -m}=\gamma_{5}\frac{1}{\not \ell -m}+\frac{1}{\not \ell +\not k
-m}\gamma_{5} \label{canc}%
\end{equation}
The above identity valid at $n=4$ is derived by decomposing the vertex factor
$\left(  \not k  -2m\right)  \gamma_{5}$ into $\left(  \not \ell +\not k
-m\right)  \gamma_{5}$ and $\gamma_{5}\left(  \not \ell -m\right)  $ to
annihilate respectively the propagators of the outgoing fermion with momentum
$\ell+k$ and the incoming fermion with momentum $\ell$. Positioning
$\gamma_{5}$ at the rightmost site, $\left(  \ref{canc}\right)  $ becomes%
\begin{equation}
\frac{1}{\not \ell +\not k  -m}\left(  \not k  -2m\right)  \frac
{1}{-\not \ell -m}\gamma_{5}=\left(  \frac{1}{-\not \ell -m}+\frac
{1}{\not \ell +\not k  -m}\right)  \gamma_{5} \label{canc1}%
\end{equation}
If we disregard the rightmost $\gamma_{5}$ on both sides of the above
identity, we obtain another identity%
\begin{equation}
\frac{1}{\not \ell +\not k  -m}\left(  \not k  -2m\right)  \frac
{1}{-\not \ell -m}=\frac{1}{-\not \ell -m}+\frac{1}{\not \ell +\not k  -m}
\label{canc2}%
\end{equation}
that is valid at $n=4$. This new identity $\left(  \ref{canc2}\right)  $,
which is void of $\gamma_{5}$, may be analytically continued to hold when
$n\neq4$. We then multiply $\gamma_{5}$ on the right to every analytically
continued term of this $\gamma_{5}$-free identity $\left(  \ref{canc2}\right)
$ to yield the analytic continuation of the identity $\left(  \ref{canc}%
\right)  $.

As a side remark, we note that when we go to the dimension of $n\neq4$,
$\left(  \ref{canc}\right)  $ in the form presented above is not valid. This
is because $\gamma_{5}$ does not always anti-commute with $\gamma^{\mu}$ if
$n\neq4$. Adopting the rightmost $\gamma_{5}$ ordering avoids this difficulty,
as the validity of the identity in the form of rightmost $\gamma_{5}$ ordering
no longer depends on $\gamma_{5}$ anti-commuting with the $\gamma$ matrices.

Before analytic continuation is made, a $\gamma_{5}$-odd ($\gamma_{5}$-even)
matrix product may always be reduced to a matrix product with only one (zero)
$\gamma_{5}$ factor. For an amplitude corresponding to a diagram involving no
fermion loops, we shall move all $\gamma_{5}$ matrices to the rightmost
position before we continue analytically the dimension $n$. Subsequent
application of dimensional regularization gives us regulated amplitudes
satisfying the Ward identities. An identity relating the traces of matrix
products without $\gamma_{5}$ at $n=4$ can always be analytically continued to
hold when $n\neq4$. Therefore, the portion of an amplitude in which the count
of $\gamma_{5}$ on every fermion loop is even has no $\gamma_{5}$ difficulty
and does not violate any Ward identity in this continuation scheme.

For any 1-loop Feynman diagram, the amplitude calculated according to the
rightmost $\gamma_{5}$ scheme can be easily compared to that calculated
according to the BM scheme. In fact, the difference between these two
amplitudes can be straightforwardly calculated. If the rightmost $\gamma_{5}$
scheme is a gauge invariant scheme, we should be able to attribute the
difference to the amplitude due to local counter terms that are required to
restore BRST\ symmetry. It will be verified below with detailed results that
this is indeed what happens. For the chiral Abelian-Higgs theory, the finite
counter terms obtained by calculating the difference between the rightmost
$\gamma_{5}$ scheme and the BM scheme are found to be exactly the same as
those obtained in \cite{SANR}. For the chiral non-Abelian gauge theory, the
finite counter terms obtained in \cite{MS} can also be accounted for by the
difference as demonstrated in Sec. \ref{n1r}. These results serve to confirm
that the rightmost $\gamma_{5}$ scheme is indeed a gauge invariant
regularization scheme.

\section{Lagrangian for the Chiral Abelian-Higgs Theory}

The BRST \cite{BRST} invariant Lagrangian density for the chiral Abelian-Higgs
gauge theory is%
\begin{align}
L_{B}  &  =-{\frac{1}{4}}F_{\mu\nu}F^{\mu\nu}+\left(  D^{\mu}\phi\right)
^{\dagger}\left(  D_{\mu}\phi\right)  -\frac{1}{2}\lambda g^{2}\left(
\phi^{\dagger}\phi-\frac{1}{2}v^{2}\right)  ^{2}\label{e2-1}\\
&  +\bar{\psi}_{L}\left(  i\not D  \right)  \psi_{L}+\bar{\psi}_{R}\left(
i\not \partial \right)  \psi_{R}-\sqrt{2}f\left(  \bar{\psi}_{L}\phi\psi
_{R}+\bar{\psi}_{R}\phi^{\dagger}\psi_{L}\right) \nonumber\\
&  -\frac{1}{2\alpha}\left(  \partial_{\mu}A^{\mu}-\alpha\Lambda\phi
_{2}\right)  ^{2}+i\bar{c}\left(  \partial_{\mu}\partial^{\mu}+\alpha\Lambda
M\right)  c+ig\alpha\Lambda\bar{c}Hc\nonumber
\end{align}
where $c$ is the ghost field, $\bar{c}$ is the anti-ghost field, and%
\[
F_{\mu\nu}\equiv\partial_{\mu}A_{\nu}-\partial_{\nu}A_{\mu},
\]%
\[
D_{\mu}\phi\equiv\left(  \partial_{\mu}+igA_{\mu}\right)  \phi,
\]%
\[
\psi_{L}=L\psi,\,\psi_{R}=R\psi
\]
with the chiral projection operators $L$ and $R$ defined as%
\[
L=\frac{1}{2}\left(  1-\gamma_{5}\right)  ,R=\frac{1}{2}\left(  1+\gamma
_{5}\right)  .
\]
The complex scalar field $\phi$ is related to real $H$ and $\phi_{2}$ by
\begin{equation}
\phi=\dfrac{H+i\phi_{2}+v}{\sqrt{2}}. \label{defphi}%
\end{equation}
The Lagrangian $L_{B}$ of $\left(  \ref{e2-1}\right)  $ is invariant under the
BRST variations:%
\begin{align}
\delta_{B}A_{\mu}  &  =\partial_{\mu}c,\label{brst}\\
\delta_{B}\phi_{2}  &  =-Mc-gcH,\nonumber\\
\delta_{B}H  &  =gc\phi_{2},\nonumber\\
\delta_{B}\psi_{L}  &  =-igc\psi_{L},\delta_{B}\psi_{R}=0,\nonumber\\
\delta_{B}\bar{c}  &  =-\dfrac{i}{\alpha}\left(  \partial^{\mu}A_{\mu}%
-\alpha\Lambda\phi_{2}\right)  ,\delta_{B}c=0.\nonumber
\end{align}
Define two mass parameters $M$ and $m$ by%
\begin{equation}
M=gv,m=fv \label{mdef}%
\end{equation}
Both $M$ and $m$ will be regarded as zero order quantities in perturbation.

Let us introduce the notation $\underline{p^{\mu}}$ for the component of
$p^{\mu}$ vector in the first 4 dimensions and the notation $p_{\Delta}^{\mu}$
for the component in the remaining dimensions. $i.e.,$%
\[
p^{\mu}=\underline{p}^{\mu}+p_{\Delta}^{\mu},
\]
with%
\[
p_{\Delta}^{\mu}=0\text{\ if }\mu\in\left\{  0,1,2,3\right\}  ,\text{
}\underline{p}^{\mu}=0\text{\ if }\mu\notin\left\{  0,1,2,3\right\}  .
\]
Likewise, the Dirac matrix $\gamma^{\mu}$ is decomposed as%
\[
\gamma^{\mu}=\underline{\gamma}^{\mu}+\gamma_{\Delta}^{\mu}%
\]
with $\gamma_{\Delta}^{\mu}=0$\ when $\mu\in\left\{  0,1,2,3\right\}  $ and
$\underline{\gamma}^{\mu}=0$ when $\mu\notin\left\{  0,1,2,3\right\}  $. Since
the definition $\left(  \ref{g5def}\right)  $ for $\gamma_{5}$ is valid even
when the space-time dimension $n$ departs from $4$, we have
\begin{equation}
\gamma_{5}\gamma^{\mu}+\gamma^{\mu}\gamma_{5}=2\gamma_{\Delta}^{\mu}\gamma
_{5}. \label{g5ant2}%
\end{equation}

The free term involving the fermion fields in $\left(  \ref{e2-1}\right)  $ is
equal to%
\begin{align}
&  \bar{\psi}_{L}i\not \partial \psi_{L}+\bar{\psi}_{R}i\not \partial \psi
_{R}-fv\left(  \bar{\psi}_{L}\psi_{R}+\bar{\psi}_{R}\psi_{L}\right)
\label{L0f}\\
&  =\bar{\psi}\left(  iR\not \partial L+iL\not \partial R-m\right)
\psi\nonumber\\
&  =\bar{\psi}\left(  i\underline{\not \partial }-m\right)  \psi\nonumber
\end{align}
where%
\[
\underline{\not \partial }=\partial_{\mu}\underline{\gamma}^{\mu}%
=\partial_{\mu}\gamma^{\mu}-\partial_{\mu}\gamma_{\Delta}^{\mu}=\not \partial
-\not \partial _{\Delta}%
\]
The fermion propagator corresponding to the free Lagrangian $\left(
\ref{L0f}\right)  $ is
\[
\frac{i}{\underline{\not p }-m}%
\]
which is independent of $p_{\Delta}$, the component of the momentum $p$ in the
extra-4 dimensions and cannot be used for perturbative dimensional calculation.

To remedy this ill behavior, let us add the term%
\begin{equation}
E_{0}=\bar{\psi}i\not \partial _{\Delta}\psi=\bar{\psi}_{R}i\not \partial
\psi_{L}+\bar{\psi}_{L}i\not \partial \psi_{R} \label{d0}%
\end{equation}
to the BRST invariant $L_{B}$ of $\left(  \ref{e2-1}\right)  $. The theory
defined by the Lagrangian%
\begin{equation}
L_{eff}=L_{B}+E_{0} \label{Lbm}%
\end{equation}
will have well-behaved free fermion propagator%
\[
\frac{i}{\not p  -m}%
\]
and can be used to calculate amplitudes perturbatively under the BM
dimensional regularization scheme. By doing so, we also incur a loss of the
BRST symmetry since $\delta_{B}L_{eff}=\delta_{B}E_{0}\neq0$. Because $E_{0}$
vanishes as $n\rightarrow4$, $E_{0}$ does not have any tree-level
contribution. At one or more loop orders, simple $\frac{1}{n-4}$ pole factors
or higher pole terms may arise from divergent loop integrals so that the
contribution of $E_{0}$ cannot be neglected and additional local counter terms
are required to restore the BRST symmetry.

For the Abelian theory with the Lagrangian $\left(  \ref{Lbm}\right)  $, the
propagators can be readily read off the free Lagrangian and vertex factors can
be determined from the interaction terms in the Lagrangian. The propagators
and vertices that are relevant to 1-loop finite counter-term calculation are
listed in Appendix \ref{frs}.

\section{Difference between the Rightmost $\gamma_{5}$ Scheme and the BM
Scheme in the Abelian-Higgs theory.\label{exsc}}

To illustrate how the counter-term amplitude is evaluated by calculating the
difference between the rightmost $\gamma_{5}$ scheme and the BM scheme for the
chiral Abelian-Higgs theory, consider the fermion self-energy diagram:%
\begin{equation}%
\begin{picture}(92,25) (2,-18)
\SetWidth{0.5}
\SetColor{Black}
\PhotonArc[clock](50,-15)(20,-180,-360){1.5}{8.5}
\SetWidth{1.0}
\Line
[arrow,arrowpos=0.5,arrowlength=3,arrowwidth=1.2,arrowinset=0.2](70,-15)(30,-15)
\Text(17,-6)[]{\normalsize{\Black{$A$}}}
\Text(77,-5)[]{\normalsize{\Black{$A$}}}
\end{picture}
\label{bm-fct1}%
\end{equation}
The horizontal line signifies an internal fermion line and the wavy line is a
vector meson line. The Feynman amplitude in the BM scheme is%
\begin{align}
\Gamma^{BM}  &  =\left(  -ig\right)  ^{2}\int\frac{d^{n}\ell}{\left(
2\pi\right)  ^{n}}D\left(  A^{\mu},A^{\nu};\ell\right)  R\gamma^{\mu}L\frac
{i}{\not \ell +\not p  -m}R\gamma^{\nu}L\label{fc1}\\
&  =-ig^{2}\int\frac{d^{n}\ell}{\left(  2\pi\right)  ^{n}}D\left(  A^{\mu
},A^{\nu};\ell\right)  \frac{R\gamma^{\mu}L\left(  \not \ell +\not p  \right)
R\gamma^{\nu}L}{\left(  \ell+p\right)  ^{2}-m^{2}}\nonumber
\end{align}
where $\ell$ is the momentum of the internal vector meson line and the
external momentum $p$ flowing into the fermion self-energy correction has only
components in the first 4 dimensions. Anti-commuting $\gamma_{5}$ to the
rightmost position, we obtain the corresponding amplitude in the rightmost
$\gamma_{5}$ scheme:%
\begin{equation}
\Gamma^{R5}=-ig^{2}\int\frac{d^{n}\ell}{\left(  2\pi\right)  ^{n}}D\left(
A^{\mu},A^{\nu};\ell\right)  \frac{\gamma^{\mu}\left(  \not \ell +\not p
\right)  \gamma^{\nu}L}{\left(  \ell+p\right)  ^{2}-m^{2}} \label{fc2}%
\end{equation}
Both $\left(  \ref{fc1}\right)  $ and $\left(  \ref{fc2}\right)  $ are
linearly divergent. Since the difference $\left(  \Gamma^{R5}-\Gamma
^{BM}\right)  $ contains at least a factor of $\gamma_{\Delta}$ matrix in the
extra-4 dimensions, terms that are convergent will not survive
the$\ n\rightarrow4$ limit. We are free to change the mass pole of any
propagator in evaluating $\left(  \Gamma^{R5}-\Gamma^{BM}\right)  $ because
the terms neglected are proportional to the mass square difference and are
therefore convergent by power counting. Furthermore, if we expand the
amplitude in a Taylor series with respect the external momentum $p$, terms
proportional to $p^{N}$ with $N\geq2$ are convergent and can be discarded in
the difference between $\left(  \ref{fc1}\right)  $ and $\left(
\ref{fc2}\right)  $. i.e., we may substitute $-i\frac{\left(  g^{\mu\nu
}+\left(  \alpha-1\right)  \frac{\ell^{\mu}\ell^{\nu}}{\ell^{2}-m^{2}}\right)
}{\left(  \ell^{2}-m^{2}\right)  ^{2}}$ for $D\left(  A^{\mu},A^{\nu}%
;\ell\right)  $ and $\frac{1}{\ell^{2}-m^{2}}\left(  1-\frac{2\ell\cdot
p}{\ell^{2}-m^{2}}\right)  $ for $\frac{1}{\left(  \ell+p\right)  ^{2}-m^{2}}%
$. The difference $\left(  \Gamma^{R5}-\Gamma^{BM}\right)  $ after utilizing
$R\gamma^{\mu}L=\underline{\gamma}^{\mu}L,R\gamma^{\nu}L=\underline{\gamma
}^{\nu}L$ and $L\not \ell R=\underline{\not \ell }R$ can be written$\ $as%
\begin{align}
\lim_{n\rightarrow4}\left(  \Gamma^{R5}-\Gamma^{BM}\right)   &  =-g^{2}%
\lim_{n\rightarrow4}\int\frac{d^{n}\ell}{\left(  2\pi\right)  ^{n}}%
\frac{\left(  g^{\mu\nu}+\left(  \alpha-1\right)  \frac{\ell^{\mu}\ell^{\nu}%
}{\ell^{2}-m^{2}}\right)  \left(  1-\frac{2\ell\cdot p}{\ell^{2}-m^{2}%
}\right)  }{\left(  \ell^{2}-m^{2}\right)  ^{2}}\label{fc3}\\
&  \times\left(  \gamma^{\mu}\left(  \not \ell +\not p  \right)  \gamma^{\nu
}-\underline{\gamma}^{\mu}\left(  \underline{\not \ell }+\not p  \right)
\underline{\gamma}^{\nu}\right)  L\nonumber
\end{align}
The symmetric integrals%
\begin{align*}
\int d^{n}\ell f\left(  \ell^{2}\right)  \ell^{\mu}\ell^{\nu}  &
=\frac{g^{\mu\nu}}{n}\int d^{n}\ell f\left(  \ell^{2}\right)  \ell^{2},\\
\int d^{n}\ell f\left(  \ell^{2}\right)  \ell^{\mu}\ell^{\nu}\ell^{\rho}%
\ell^{\sigma}  &  =\frac{\left(  g^{\mu\nu}g^{\rho\sigma}+g^{\mu\rho}%
g^{\nu\sigma}+g^{\mu\sigma}g^{\rho\nu}\right)  }{n\left(  n+2\right)  }\int
d^{n}\ell f\left(  \ell^{2}\right)  \ell^{4}%
\end{align*}
enable us to set%
\begin{align*}
\int d^{n}\ell f\left(  \ell^{2}\right)  \left(  \ell\cdot p\right)
\not \ell  &  =\frac{1}{n}\int d^{n}\ell f\left(  \ell^{2}\right)  \ell
^{2}\not p \\
\int d^{n}\ell f\left(  \ell^{2}\right)  \ell^{\mu}\ell^{\nu}\left(  \ell\cdot
p\right)  \not \ell  &  =\frac{g^{\mu\nu}\not p  +p^{\mu}\gamma^{\nu}+p^{\nu
}\gamma^{\mu}}{n\left(  n+2\right)  }\int d^{n}\ell f\left(  \ell^{2}\right)
\ell^{4}%
\end{align*}
and reduce $\left(  \ref{fc3}\right)  $ to%
\begin{align}
\lim_{n\rightarrow4}\left(  \Gamma^{R5}-\Gamma^{BM}\right)   &  =\frac{g^{2}%
}{6}\lim_{n\rightarrow4}\int\frac{d^{n}\ell}{\left(  2\pi\right)  ^{n}}%
\frac{\left(  n-4\right)  }{\left(  \ell^{2}-m^{2}\right)  ^{2}}\left(
1+2\alpha\right)  \not p  L\label{fc4}\\
&  =-\frac{1}{\left(  4\pi\right)  ^{2}}i\frac{g^{2}}{3}\left(  1+2\alpha
\right)  \not p  L\nonumber
\end{align}
where we have also utilized the integral
\[
\lim_{n\rightarrow4}\int\frac{d^{n}\ell}{\left(  2\pi\right)  ^{n}}%
\frac{\left(  n-4\right)  }{\left(  \ell^{2}-m\right)  ^{2}}=4\int\frac
{d^{4}\ell}{\left(  2\pi\right)  ^{4}}\frac{m^{2}}{\left(  \ell^{2}-1\right)
^{3}}=\frac{-2i}{\left(  4\pi\right)  ^{2}}%
\]
In the BM scheme, this amplitude $\left(  \ref{fc4}\right)  $ can be accounted
for by adding the counter term
\begin{equation}
-\frac{1}{\left(  4\pi\right)  ^{2}}\frac{g^{2}}{3}\left(  1+2\alpha\right)
\bar{\psi}Ri\not \partial L\psi\label{fc5}%
\end{equation}
to the Lagrangian $\left(  \ref{Lbm}\right)  $.

The counter-term amplitude for any divergent 1-loop 1PI diagram can be
similarly calculated. The diagrams that are responsible for all 1-loop counter
terms are listed in Figure \ref{bm-o2} - \ref{bm-c4} in Appendix \ref{apxa}.
The corresponding counter-term amplitudes are calculated and summarized in
Table \ref{ct-o2} - \ref{ct-c4}.

\section{1-Loop Results for the Chiral Abelian-Higgs Theory}

For the chiral Abelian-Higgs theory, D. Sanchez-Ruiz \cite{SANR} has
successfully computed in the BM scheme the 1-loop finite counter terms that
are required to restore the BRST symmetry by evaluating the terms that break
the Slavnov-Taylor identities and then solving a linear system of 27 equations
with 32 variables to find a solution. The calculation in \cite{SANR} is rather
cumbersome and has to be relied on computer routines. The general solution for
the 1-loop finite counter terms is given by (30) of \cite{SANR}
\begin{equation}
\hbar\tilde{S}_{fct}^{\left(  1\right)  }=-\sum_{i=1}^{32}\hbar\tilde{x}%
_{0,i}^{\left(  1\right)  }\tilde{e}_{i}+\hbar\sum_{l=1}^{11}c_{l}^{\left(
1\right)  }\mathcal{I}_{l} \label{rc0}%
\end{equation}
where each $\mathcal{I}_{l}$, $i=1,2..,11$ is BRST invariant and $\tilde
{e}_{i}$, $i=1,2..,32$ given by $\left(  16\right)  $ in \cite{SANR}\ form a
basis of the space of the integrated Lorentz scalar CP-invariant polynomials
in the fields and their derivatives with maximal canonical dimension 4 and
ghost number 0. A particular solution for the coefficients $\tilde{x}%
_{0,i}^{\left(  1\right)  }$ is given by $\left(  29\right)  $ in \cite{SANR}
and tabulated in the 2nd column of Table \ref{tbcmp_open}, \ref{tbcmp_closed}
and \ref{tbcmp_closed2}. Counter terms that involve fermion fields are listed
in Table \ref{tbcmp_open}. Otherwise, they are listed in Table
\ref{tbcmp_closed} and \ref{tbcmp_closed2}.
\begin{table}[htp] \centering
\begin{tabular}
[c]{|l|l|l|l|}\hline
$\tilde{e}_{i}$ & $-\left(  4\pi\right)  ^{2}\tilde{x}_{0,i}^{\left(
1\right)  }$ & $%
\begin{array}
[c]{c}%
\xi^{\prime}=\alpha,\rho=-\alpha\Lambda,\\
\theta=0,r=1
\end{array}
$ & $%
\begin{array}
[c]{c}%
\text{rightmost }\gamma_{5}\\
\text{method}%
\end{array}
$\\\hline
$\tilde{e}_{24}=\bar{\psi}\psi$ & $-\frac{f\left[  3\rho r+4g^{2}\theta\left(
1+\theta r\right)  v\left(  5+\xi^{\prime}\right)  \right]  }{6r}$ & $\frac
{1}{2}\alpha\Lambda f$ & $-\frac{1}{2}\alpha\Lambda fg$\\
$\tilde{e}_{25}=\bar{\psi}i\not \partial L\psi$ & $0$ & $0$ & $0$\\
$\tilde{e}_{26}=\bar{\psi}i\not \partial R\psi$ & $0$ & $0$ & $0$\\
$\tilde{e}_{27}=\bar{\psi}\not A  L\psi$ & $-\frac{\left[  -6f^{2}%
r+g^{2}\left(  2\theta+r+\theta^{2}r\right)  \left(  5+\xi^{\prime}\right)
\right]  }{6}$ & $f^{2}-\frac{1}{6}g^{2}\left(  5+\alpha\right)  $ &
$-f^{2}g+\frac{1}{6}g^{3}\left(  5+\alpha\right)  $\\
$\tilde{e}_{28}=\bar{\psi}\not A  R\psi$ & $\frac{r\left[  -6f^{2}+g^{2}%
\theta^{2}\left(  5+\xi^{\prime}\right)  \right]  }{6}$ & $-f^{2}$ & $f^{2}%
g$\\
$\tilde{e}_{29}=\bar{\psi}H\psi$ & $-\frac{2}{3}fg^{2}\theta\left(
\theta+r\right)  \left(  5+\xi^{\prime}\right)  $ & $0$ & $0$\\
$\tilde{e}_{30}=\bar{\psi}\phi_{2}\gamma_{5}\psi$ & $0$ & $0$ & $0$\\\hline
\end{tabular}
\caption{Counter terms due to diagrams with open fermion lines\label{tbcmp_open}}\label{tbr1}%
\end{table}%

The theory defined by the $L_{eff}$ of $\left(  \ref{Lbm}\right)  $
corresponds to the theory of $\left(  5\right)  $ in \cite{SANR} with
$\xi^{\prime}=\alpha,\rho=-\alpha\Lambda,\theta=0$ and $r=1$. Correspondingly,
the 3rd column of Table \ref{tbcmp_open} is obtained from the 2nd column with
the substitution $\xi^{\prime}=\alpha,\rho=-\alpha\Lambda,\theta=0$ and $r=1$.

From Table \ref{ct-o2}, \ref{ct-o3A}, \ref{ct-o3H} and \ref{ct-o3phi2}, the
sum of counter terms due to the diagrams from Figure \ref{bm-o2},
\ref{bm-o3A}, \ref{bm-o3H}, and \ref{bm-o3phi2} obtained according to the
method of evaluating the difference between the rightmost $\gamma_{5}$ scheme
and the BM scheme is equal to%
\begin{equation}
\frac{1}{\left(  4\pi\right)  ^{2}}\left[
\begin{array}
[c]{c}%
-\frac{g^{2}}{3}\left(  1+2\alpha\right)  \bar{\psi}Ri\not \partial
L\psi+\frac{1}{2}\alpha\left(  M-\Lambda\right)  fg\bar{\psi}\psi\\
\frac{1}{6}g^{3}\left(  7+5\alpha\right)  \bar{\psi}R\not A  L\psi+f^{2}%
g\bar{\psi}\not A  \gamma_{5}\psi+\frac{\alpha g^{2}f}{2}\bar{\psi}\left(
H+i\phi_{2}\gamma_{5}\right)  \psi
\end{array}
\right]  \label{rc1}%
\end{equation}
which after subtracting out the gauge invariant term%
\[
\frac{1}{\left(  4\pi\right)  ^{2}}\left[  -\frac{g^{2}}{3}\left(
1+2\alpha\right)  \bar{\psi}R\left(  i\not \partial -g\not A  \right)
L\psi+\frac{\alpha g^{2}f}{2}\bar{\psi}\left(  H+v+i\phi_{2}\gamma_{5}\right)
\psi\right]
\]
becomes%
\begin{equation}
\frac{1}{\left(  4\pi\right)  ^{2}}\left[  -\frac{1}{2}\alpha\Lambda
fg\bar{\psi}\psi+\left(  \frac{1}{6}g^{3}\left(  5+\alpha\right)
-f^{2}g\right)  \bar{\psi}R\not A  L\psi+f^{2}g\bar{\psi}\not A  R\psi\right]
. \label{rc2}%
\end{equation}
The above expression $\left(  \ref{rc2}\right)  $ decomposed as a linear
combination of $\tilde{e}_{i}$ is listed under the column "rightmost
$\gamma_{5}$ method" of Table \ref{tbcmp_open}. The Lagrangian $\left(
\ref{Lbm}\right)  $ is defined differently from that in \cite{SANR}. The
covariant derivative $D_{\mu}$ is defined as $D_{\mu}=\partial_{\mu}+igA_{\mu
}$ for the theory $\left(  \ref{Lbm}\right)  $ but as $D_{\mu}=\partial_{\mu
}-iA_{\mu}$ in \cite{SANR}. The vector field $A$ in a counter-term expression
obtained in \cite{SANR} needs to be scaled to $-gA$ to be identified as the
corresponding counter term for the theory of $\left(  \ref{Lbm}\right)  $.
There is a single vector $A$ field in $\tilde{e}_{27}=\bar{\psi}\not A  L\psi$
or $\tilde{e}_{28}=\bar{\psi}\not A  R\psi$. As a consequence, multiplying
$-g$ to the coefficient of either $\tilde{e}_{27}$ or $\tilde{e}_{28}$ under
the 3rd column of Table \ref{tbcmp_open} should give us the corresponding
coefficient under the column "rightmost $\gamma_{5}$ method". Furthermore, the
counter term proportional to $\tilde{e}_{24}=\bar{\psi}\psi$ actually stems
from diagram $\left(  d\right)  $ and $\left(  e\right)  $ of Figure
$\ref{bm-o2}$. The ratio of the coefficient of $\tilde{e}_{24}$ for the theory
of $\left(  \ref{Lbm}\right)  $ over that obtained in \cite{SANR} as shown in
Table \ref{tbcmp_open} is equal to $-g$ and can be accounted for by the ratio
of vertex factors due to the single $\bar{\psi}-A-\psi$ vertex in each of the
diagram $\left(  d\right)  $ and $\left(  e\right)  $ of Figure $\ref{bm-o2}$.
Table \ref{tbcmp_open} in fact shows that the 1-loop counter terms that
involve fermion fields calculated by the rightmost $\gamma_{5}$ method are in
agreement with those obtained in \cite{SANR}.

The 1-loop counter terms due to the diagrams without external fermion lines
from Figure \ref{bm-c2}, \ref{bm-c3} and \ref{bm-c4} are summarized in Table
\ref{ct-c2}, \ref{ct-c3} and \ref{ct-c4}. The total of these counter terms is%
\begin{equation}
\frac{1}{\left(  4\pi\right)  ^{2}}\left[
\begin{array}
[c]{c}%
4m^{2}f^{2}\left(  \phi_{2}\right)  ^{2}-\frac{4}{3}f^{2}\left(  \partial
\phi_{2}\right)  ^{2}-g^{2}m^{2}A^{2}+\frac{1}{6}g^{2}\left(  \partial_{\mu
}A_{\nu}\right)  \left(  \partial^{\mu}A^{\nu}\right) \\
+8f^{3}mH\left(  \phi_{2}\right)  ^{2}-2fg^{2}mHA^{2}+4f^{2}g\phi_{2}\left(
\partial_{\mu}H\right)  A^{\mu}\\
+4f^{4}H^{2}\left(  \phi_{2}\right)  ^{2}+\frac{8}{3}f^{4}\left(  \phi
_{2}\right)  ^{4}-f^{2}g^{2}H^{2}A^{2}-3f^{2}g^{2}\left(  \phi_{2}\right)
^{2}A^{2}+\frac{1}{12}g^{4}\left(  A^{2}\right)  ^{2}%
\end{array}
\right]  \label{rc3}%
\end{equation}
and can be written as the sum of%
\begin{equation}
\frac{1}{\left(  4\pi\right)  ^{2}}\left[  2f^{4}\left(  \left(  H+v\right)
^{2}+\left(  \phi_{2}\right)  ^{2}\right)  ^{2}-2f^{4}v^{4}+\frac{1}{12}%
g^{2}F_{\mu\nu}F^{\mu\nu}\right]  \label{rc4}%
\end{equation}
and%
\begin{equation}
\frac{1}{\left(  4\pi\right)  ^{2}}\left[
\begin{array}
[c]{c}%
-2f^{4}\left(  H^{4}+4vH^{3}+6v^{2}H^{2}+4v^{3}H\right)  +\frac{2}{3}%
f^{4}\left(  \phi_{2}\right)  ^{4}-\frac{2}{3}f^{2}\left(  \partial\phi
_{2}\right)  ^{2}\\
+\frac{1}{6}g^{2}\left(  \partial_{\mu}A^{\mu}\right)  ^{2}-g^{2}m^{2}%
A^{2}-2fg^{2}mHA^{2}+4f^{2}g\phi_{2}\left(  \partial_{\mu}H\right)  A^{\mu}\\
-f^{2}g^{2}H^{2}A^{2}+\frac{1}{12}g^{4}\left(  A^{2}\right)  ^{2}-3f^{2}%
g^{2}\left(  \phi_{2}\right)  ^{2}A^{2}%
\end{array}
\right]  . \label{rc5}%
\end{equation}
$\left(  \ref{rc4}\right)  $ is gauge invariant and $\left(  \ref{rc5}\right)
$ expressed as a linear combination of $\tilde{e}_{i}$ is tabulated in Table
\ref{tbcmp_closed} and \ref{tbcmp_closed2} under the column "rightmost
$\gamma_{5}$ method" while the result from \cite{SANR} is listed under the
column $-\left(  4\pi\right)  ^{2}\tilde{x}_{0,i}^{\left(  1\right)  }$.
Taking into the consideration of the $-g$ factor for each vector field $A$ in
comparing the counter-term expressions, the counter terms listed in Table
\ref{tbcmp_closed} and \ref{tbcmp_closed2} obtained by the rightmost
$\gamma_{5}$ method for the theory of $\left(  \ref{Lbm}\right)  $ are in
exact agreement with those obtained in \cite{SANR}.%

\begin{table}[htp] \centering
\begin{tabular}
[c]{|l|l|l|}\hline
$\tilde{e}_{i}$ & $-\left(  4\pi\right)  ^{2}\tilde{x}_{0,i}^{\left(
1\right)  }$ & rightmost $\gamma_{5}$ method\\\hline
$\tilde{e}_{1}=H$ & $-8f^{4}v^{3}$ & $-8f^{4}v^{3}$\\
$\tilde{e}_{2}=H^{2}$ & $-12f^{4}v^{2}$ & $-12f^{4}v^{2}$\\
$\tilde{e}_{3}=\left(  \phi_{2}\right)  ^{2}$ & $0$ & $0$\\
$\tilde{e}_{4}=H^{3}$ & $-8f^{4}v$ & $-8f^{4}v$\\
$\tilde{e}_{5}=H\left(  \phi_{2}\right)  ^{2}$ & $0$ & $0$\\
$\tilde{e}_{6}=H^{4}$ & $-2f^{4}$ & $-2f^{4}$\\
$\tilde{e}_{7}=\left(  \phi_{2}\right)  ^{4}$ & $\frac{2}{3}f^{4}$ & $\frac
{2}{3}f^{4}$\\
$\tilde{e}_{8}=H^{2}\left(  \phi_{2}\right)  ^{2}$ & $0$ & $0$\\
$\tilde{e}_{9}=\left(  \partial_{\mu}H\right)  \left(  \partial^{\mu}H\right)
$ & $0$ & $0$\\
$\tilde{e}_{10}=\left(  \partial_{\mu}\phi_{2}\right)  \left(  \partial^{\mu
}\phi_{2}\right)  $ & $-\frac{2}{3}f^{2}$ & $-\frac{2}{3}f^{2}$\\\hline
\end{tabular}
\caption{Counter terms without A fields due to diagrams with a closed fermion loop\label{tbcmp_closed}}\label{tbr2}%
\end{table}%

\begin{table}[htp] \centering
\begin{tabular}
[c]{|l|l|l|}\hline
$\tilde{e}_{i}$ & $-\left(  4\pi\right)  ^{2}\tilde{x}_{0,i}^{\left(
1\right)  }$ & rightmost $\gamma_{5}$ method\\\hline
$\tilde{e}_{11}=\phi_{2}\left(  \partial_{\mu}A^{\mu}\right)  $ & $0$ & $0$\\
$\tilde{e}_{12}=A_{\mu}H\left(  \partial^{\mu}\phi_{2}\right)  $ & $0$ & $0$\\
$\tilde{e}_{13}=A_{\mu}\phi_{2}\left(  \partial^{\mu}H\right)  $ & $-4f^{2}$ &
$4f^{2}g$\\
$\tilde{e}_{14}=A_{\mu}A^{\mu}$ & $-f^{2}v^{2}$ & $-f^{2}g^{2}v^{2}$\\
$\tilde{e}_{15}=A_{\mu}A^{\mu}H$ & $-2f^{2}v$ & $-2f^{2}g^{2}v$\\
$\tilde{e}_{16}=A_{\mu}A^{\mu}H^{2}$ & $-f^{2}$ & $-f^{2}g^{2}$\\
$\tilde{e}_{17}=A_{\mu}A^{\mu}\left(  \phi_{2}\right)  ^{2}$ & $-3f^{2}$ &
$-3f^{2}g^{2}$\\
$\tilde{e}_{18}=\left(  \partial_{\mu}A^{\mu}\right)  ^{2}$ & $\frac{1}{6}$ &
$\frac{1}{6}g^{2}$\\
$\tilde{e}_{19}=F_{\mu\nu}F^{\mu\nu}$ & $0$ & $0$\\
$\tilde{e}_{20}=\left(  A_{\mu}A^{\mu}\right)  ^{2}$ & $\frac{1}{12}$ &
$\frac{1}{12}g^{4}$\\\hline
\end{tabular}
\caption{Counter terms involving A fields due to diagrams with a closed fermion loop\label{tbcmp_closed2}}\label{tbr3}%
\end{table}%

\section{Lagrangian for the Chiral Non-Abelian Gauge Theory}

The BRST invariant Lagrangian density for the chiral non-Abelian gauge theory
is%
\begin{align}
\tilde{L}_{B}  &  =-\frac{1}{4}F_{\mu\nu}^{a}F^{a,\mu\nu}+\bar{\psi}%
_{L}i\not D \psi_{L}+\bar{\psi}_{R}i\not \partial \psi_{R}+i\bar{\psi}%
_{R}^{\prime}i\not D \psi_{R}^{\prime}+\bar{\psi}_{L}^{\prime}i\not \partial
\psi_{L}^{\prime}\label{Lnab}\\
&  -\frac{1}{2\alpha}\left(  \partial^{\mu}A_{\mu}^{a}\right)  \left(
\partial^{\nu}A_{\nu}^{a}\right)  +i\bar{c}^{a}\square c^{a}+igC^{abc}\left(
\partial^{\mu}\bar{c}^{a}\right)  A_{\mu}^{b}c^{c}\nonumber
\end{align}
where $c^{a}$ is the ghost field, $\bar{c}^{a}$ is the anti-ghost field, and
\[
F_{\mu\nu}^{a}=\partial_{\mu}A_{\nu}^{a}-\partial_{\nu}A_{\mu}^{a}%
-gC^{abc}A_{\mu}^{b}A_{\nu}^{c}%
\]
Two fermion fields $\psi$ and $\psi^{\prime}$ whose left-handed component
$\psi_{L}=L\psi$ and right-handed component $\psi_{R}^{\prime}=R\psi^{\prime}$
are coupled to $A_{\mu}^{a}$. The covariant derivatives for $\psi_{L}$ and
$\psi_{R}^{\prime}$ are
\begin{align*}
D_{\mu}\psi_{L}  &  =\left(  \partial_{\mu}+igA_{\mu}^{a}T_{L}^{a}\right)
\psi_{L}\\
D_{\mu}\psi_{R}^{\prime}  &  =\left(  \partial_{\mu}+igA_{\mu}^{a}T_{R}%
^{a}\right)  \psi_{R}^{\prime}%
\end{align*}
where $T_{L}^{a}$ $\left(  T_{R}^{a}\right)  $ are group generators that
satisfy
\begin{align*}
\left[  T_{L}^{a},T_{L}^{b}\right]   &  =iC^{abc}T_{L}^{c},\left[  T_{R}%
^{a},T_{R}^{b}\right]  =iC^{abc}T_{R}^{c}\\
tr\left(  T_{L}^{a}T_{L}^{b}\right)   &  =T_{L}\delta^{ab},tr\left(  T_{R}%
^{a}T_{R}^{b}\right)  =T_{R}\delta^{ab}\\
\sum_{e}T_{L}^{e}T_{L}^{e}  &  =C_{L},\sum_{e}T_{R}^{e}T_{R}^{e}=C_{R}%
\end{align*}
For convenience, we also adopt the following shorthand notations defined in
\cite{MS}:%
\[
tr\left(  T_{L}^{a}T_{L}^{b}T_{L}^{c}T_{L}^{d}\right)  =T_{L}^{abcd},tr\left(
T_{R}^{a}T_{R}^{b}T_{R}^{c}T_{R}^{d}\right)  =T_{R}^{abcd}%
\]
The Lagrangian $\left(  \ref{Lnab}\right)  $ is invariant under the BRST
variations:%
\begin{align*}
\delta_{B}A_{\mu}^{a}  &  =\partial_{\mu}c^{a}+gC^{abc}c^{b}A_{\mu}^{c}\\
\delta_{B}\psi_{L}  &  =-ic^{a}T_{L}^{a}\psi_{L},\delta_{B}\psi_{R}=0\\
\delta_{B}\psi_{R}^{\prime}  &  =-ic^{a}T_{R}^{a}\psi_{R}^{\prime},\delta
_{B}\psi_{L}^{\prime}=0\\
\delta_{B}c^{a}  &  =\frac{1}{2}gC^{abc}c^{b}c^{c}\\
\delta_{B}\bar{c}^{a}  &  =\frac{1}{i\alpha}\partial_{\mu}A^{a,\mu}%
\end{align*}

As with the chiral Abelian-Higgs theory $\left(  \ref{Lbm}\right)  $, a gauge
variant evanescent term%
\begin{equation}
\tilde{E}_{0}=\bar{\psi}i\not \partial_{\Delta}\psi+\bar{\psi}^{\prime
}i\not \partial_{\Delta}\psi^{\prime}\label{nd0}%
\end{equation}
needs to be added the BRST invariant $\left(  \ref{Lnab}\right)  $ to define
the Lagrangian%
\begin{equation}
\tilde{L}_{eff}=\tilde{L}_{B}+\tilde{E}_{0}\label{nbm}%
\end{equation}
for the chiral non-Abelian gauge theory that can be calculated perturbatively
in the BM scheme. For the non-Abelian theory $\left(  \ref{nbm}\right)  $, the
propagators and vertices that are relevant to 1-loop finite counter-term
calculation are listed in Appendix \ref{nfrs}.

\section{1-Loop Results for the Chiral Non-Abelian Gauge Theory\label{n1r}}

C. P. Martin and D. Sanchez-Ruiz have obtained with tedious calculations the
1-loop finite counter terms that are needed for restoring BRST symmetry in the
BM dimensional regularization formalism for the chiral non-Abelian gauge
theory with the result given in (69) of \cite{MS}. In Appendix \ref{apxb}, the
1-loop counter terms for this non-Abelian theory are computed
straightforwardly by evaluating the difference of amplitudes between the
rightmost $\gamma_{5}$ scheme and the BM scheme with the results summarized in
Table \ref{ct-nab}. Specifically, diagrams in Figure \ref{bm-no2} and
\ref{bm-no3A} yield the counter terms that involve fermion fields and can be
written as%
\begin{equation}
\frac{1}{\left(  4\pi\right)  ^{2}}\left[
\begin{array}
[c]{c}%
-\frac{1}{3}g^{2}\left(  1+2\alpha\right)  \left(  \bar{\psi}_{L}%
i\not \partial\psi_{L}C_{L}+\bar{\psi}_{R}^{\prime}i\not \partial\psi
_{R}^{\prime}C_{R}\right)  \\
+\frac{1}{6}g^{3}\left(  7+5\alpha\right)  \left(  \bar{\psi}_{L}\not A%
\psi_{L}C_{L}T_{L}^{a}+\bar{\psi}_{R}^{\prime}\not A\psi_{R}^{\prime}%
C_{R}T_{R}^{a}\right)
\end{array}
\right]  \label{n1r1}%
\end{equation}
Subtracting out the gauge invariant term%
\[
\frac{1}{\left(  4\pi\right)  ^{2}}\left[  -i\frac{1}{6}g^{2}\left(
7+5\alpha\right)  \left(
\begin{array}
[c]{c}%
\bar{\psi}_{L}\left(  \not \partial+ig\not A^{a}T_{L}^{a}\right)  \psi
_{L}C_{L}T_{L}^{a}\\
+\bar{\psi}_{R}^{\prime}\left(  \not \partial+ig\not A^{a}T_{R}^{a}\right)
\psi_{R}^{\prime}C_{R}T_{R}^{a}%
\end{array}
\right)  \right]
\]
from $\left(  \ref{n1r1}\right)  $, we get%
\begin{equation}
\frac{1}{\left(  4\pi\right)  ^{2}}\left[  \left(  1+\frac{\left(
\alpha-1\right)  }{6}\right)  g^{2}\left(  \bar{\psi}_{L}i\not \partial
\psi_{L}C_{L}T_{L}^{a}+\bar{\psi}_{R}^{\prime}i\not \partial\psi_{R}^{\prime
}C_{R}T_{R}^{a}\right)  \right]  \label{n1r2}%
\end{equation}
which, after the identification of $\alpha$ with $\alpha^{\prime}$, is
consistent with the finite counter-terms (69) of \cite{MS}.

Figure \ref{bm-nc2},\ref{bm-nc3}, and \ref{bm-nc4} are responsible for the
counter terms that are free of fermion fields. From Table \ref{ct-nab}, the
sum of these counter terms is equal to%
\begin{equation}
\frac{1}{\left(  4\pi\right)  ^{2}}\left[
\begin{array}
[c]{c}%
-\frac{1}{6}g^{2}\left(  T_{L}+T_{R}\right)  A_{\mu}^{a}\square A^{a,\mu}\\
-\frac{2}{3}g^{3}\left(  T_{L}+T_{R}\right)  C^{abc}\left(  \partial^{\mu
}A_{\nu}^{a}\right)  A_{\mu}^{b}A^{c,\nu}\\
\frac{1}{12}g^{4}\left(  T_{L}^{abcd}+T_{R}^{abcd}\right)  A^{a,\mu}A_{\mu
}^{b}A^{c,\nu}A_{\nu}^{d}\\
+\frac{5}{24}g^{4}\left(  T_{L}+T_{R}\right)  C^{eab}C^{ecd}A_{\mu}^{a}A_{\nu
}^{b}A^{c,\mu}A^{d,\nu}%
\end{array}
\right]  \label{n1r3}%
\end{equation}
which can be written as the sum of the gauge invariant term%
\[
\frac{5}{24}g^{2}\left(  T_{L}+T_{R}\right)  F_{\mu\nu}^{a}F^{a,\mu\nu}%
\]
and%
\begin{equation}
\frac{1}{\left(  4\pi\right)  ^{2}}\left[
\begin{array}
[c]{c}%
\left(  T_{L}+T_{R}\right)  g^{2}\left(  \frac{5}{12}\left(  \partial
A\right)  ^{2}+\frac{1}{4}A\square A\right) \\
+\frac{\left(  T_{L}+T_{R}\right)  }{6}g^{3}C^{abc}\left(  \partial_{\mu
}A_{\nu}^{a}\right)  A_{\mu}^{b}A_{\nu}^{c}\\
+\frac{1}{12}g^{4}\left(  T_{L}^{abcd}+T_{R}^{abcd}\right)  A^{a,\mu}A_{\mu
}^{b}A^{c,\nu}A_{\nu}^{d}%
\end{array}
\right]  \label{n1r4}%
\end{equation}
Upon the scaling of $A\rightarrow-gA$, the chiral non-Abelian gauge theory
with the tree action defined by (32) in \cite{MS} becomes the non-Abelian
theory defined by $\left(  \ref{Lnab}\right)  $. Taking into the consideration
of this $A\rightarrow-gA$ scaling, $\left(  \ref{n1r4}\right)  $ is also in
agreement with (69) of \cite{MS}.

\section{Conclusion}

In the BM scheme, simply removing the pole terms from the amplitudes of 1-loop
diagrams does not yield renormalized amplitudes that satisfy Ward identities.
Instead, some finite renormalization terms have to be added. These finite
counter terms are determined from restoring the validities of Ward identities.
Implementing this finite renormalization in practical calculation is usually a
daunting task even at 1-loop order.

For the chiral Abelian-Higgs gauge theory and the chiral non-Abelian
Yang-Mills theory, we have verified that the renormalized amplitudes for all
1-loop diagrams calculated in the BM scheme with finite counter term
renormalization are equal to those obtained directly in the rightmost
$\gamma_{5}$ scheme. This means we can be spared the tedious finite
renormalization procedures if the rightmost $\gamma_{5}$ scheme is adopted.
Furthermore, since all the $\gamma_{5}$ matrices are moved to and consolidated
at a single position before continuing the dimension in our scheme, the burden
of evaluating the matrix products or trace of matrix products is considerably
less than that in the BM scheme. In our opinion, this rightmost $\gamma_{5}$
prescription is a much simpler scheme than the BM scheme for calculating
amplitudes in gauge theories involving $\gamma_{5}$.

For the rightmost $\gamma_{5}$ scheme, the prescription that leads to the
preservation of Ward identities makes no use of the specific type of gauge
theories in question. As a consequence, the rightmost $\gamma_{5}$ scheme
should be applicable for any type of chiral gauge theory, in particular, the
standard model.%

\appendix\appendixpage

\section{The Chiral Abelian-Higgs Theory\label{apxa}}

\subsection{Feynman Rules\label{frs}}

The propagators and vertices used in the 1-loop counter-term calculation for
the chiral Abelian-Higgs theory defined by $\left(  \ref{Lbm}\right)  $ are
listed below.

\subsubsection{Propagators:}

\begin{center}%
\[
S\left(  \psi,\bar{\psi};p\right)  :%
\begin{array}
[b]{c}%
\begin{picture}(62,23) (29,-10)
\SetWidth{0.5}
\SetColor{Black}
\Text(60,3)[]{\normalsize{\Black{$p$}}}
\SetWidth{1.0}
\Line
[arrow,arrowpos=0.5,arrowlength=3,arrowwidth=1.2,arrowinset=0.2](90,-7)(30,-7)
\end{picture}%
\end{array}
=\frac{i}{\not p  -m}%
\]%
\[
D\left(  A^{\mu},A^{\nu};k\right)  :%
\begin{array}
[c]{c}%
\begin{picture}(92,36) (15,-10)
\SetWidth{0.5}
\SetColor{Black}
\Photon(30,6)(90,6){2.5}{6}
\Text(60,16)[]{\normalsize{\Black{$k$}}}
\Text(30,-4)[]{\normalsize{\Black{$\mu$}}}
\Text(90,-4)[]{\normalsize{\Black{$\nu$}}}
\Line
[arrow,arrowpos=1,arrowlength=2.5,arrowwidth=1,arrowinset=0.2](50,16)(40,16)
\end{picture}%
\end{array}
=-i\left(  \frac{g^{\mu\nu}-\frac{k^{\mu}k^{\nu}}{k^{2}}}{k^{2}-M^{2}}%
+\frac{\alpha\left(  k^{2}-\alpha\Lambda^{2}\right)  k^{\mu}k^{\nu}}%
{k^{2}\left(  k^{2}-\alpha\Lambda M\right)  ^{2}}\right)
\]%
\[
D\left(  A^{\mu},\phi_{2};k\right)  :%
\begin{array}
[c]{c}%
\begin{picture}(92,36) (15,-10)
\SetWidth{0.5}
\SetColor{Black}
\Photon(30,6)(60,6){2.5}{5}
\Text(60,16)[]{\normalsize{\Black{$k$}}}
\Text(30,-4)[]{\normalsize{\Black{$\mu$}}}
\Text(90,-4)[]{\normalsize{\Black{$\phi_2$}}}
\Line
[arrow,arrowpos=1,arrowlength=2.5,arrowwidth=1,arrowinset=0.2](50,16)(40,16)
\Line[dash,dashsize=0.3](60,6)(90,6)
\end{picture}%
\end{array}
=\frac{\alpha\left(  M-\Lambda\right)  k^{\mu}}{\left(  k^{2}-\alpha\Lambda
M\right)  ^{2}}%
\]%
\[
D\left(  \phi_{2},A^{\mu};k\right)  :%
\begin{array}
[c]{c}%
\begin{picture}(92,36) (15,-10)
\SetWidth{0.5}
\SetColor{Black}
\Photon(60,6)(90,6){2.5}{5}
\Text(60,16)[]{\normalsize{\Black{$k$}}}
\Text(90,-4)[]{\normalsize{\Black{$\mu$}}}
\Text(30,-4)[]{\normalsize{\Black{$\phi_2$}}}
\Line
[arrow,arrowpos=1,arrowlength=2.5,arrowwidth=1,arrowinset=0.2](50,16)(40,16)
\Line[dash,dashsize=0.3](30,6)(60,6)
\end{picture}%
\end{array}
=-\frac{\alpha\left(  M-\Lambda\right)  k^{\mu}}{\left(  k^{2}-\alpha\Lambda
M\right)  ^{2}}%
\]%
\[
D\left(  \phi_{2},\phi_{2};k\right)  :%
\begin{array}
[c]{c}%
\begin{picture}(92,36) (15,-10)
\SetWidth{0.5}
\SetColor{Black}
\Text(60,16)[]{\normalsize{\Black{$k$}}}
\Text(90,-4)[]{\normalsize{\Black{$\phi_2$}}}
\Text(30,-4)[]{\normalsize{\Black{$\phi_2$}}}
\SetWidth{0.5}
\Line
[arrow,arrowpos=1,arrowlength=2.5,arrowwidth=1,arrowinset=0.2](50,16)(40,16)
\Line[dash,dashsize=0.3](30,6)(90,6)
\end{picture}%
\end{array}
=\frac{i\left(  k^{2}-\alpha M^{2}\right)  }{\left(  k^{2}-\alpha\Lambda
M\right)  ^{2}}%
\]%
\[
D\left(  H,H;k\right)  :%
\begin{array}
[c]{c}%
\begin{picture}(92,36) (15,-10)
\SetWidth{0.5}
\SetColor{Black}
\Text(60,16)[]{\normalsize{\Black{$k$}}}
\Text(90,-4)[]{\normalsize{\Black{$H$}}}
\Text(30,-4)[]{\normalsize{\Black{$H$}}}
\SetWidth{0.5}
\Line
[arrow,arrowpos=1,arrowlength=2.5,arrowwidth=1,arrowinset=0.2](50,16)(40,16)
\Line[dash,dashsize=0.3](30,6)(90,6)
\end{picture}%
\end{array}
=\frac{i}{k^{2}-\lambda M^{2}}%
\]

\end{center}

\subsubsection{Vertex Factors:}%

\[
\bar{\psi}A^{\mu}\psi:%
\begin{array}
[c]{c}%
\begin{picture}(62,24) (59,-19)
\SetWidth{1.0}
\SetColor{Black}
\Line
[arrow,arrowpos=0.5,arrowlength=3,arrowwidth=1.2,arrowinset=0.2](120,-16)(90,-16)
\SetWidth{0.5}
\Photon(90,4)(90,-16){2.5}{4}
\SetWidth{1.0}
\Line
[arrow,arrowpos=0.5,arrowlength=3,arrowwidth=1.2,arrowinset=0.2](90,-16)(60,-16)
\Text(100,-6)[]{\normalsize{\Black{$\mu$}}}
\end{picture}%
\end{array}
=-igR\gamma^{\mu}L
\]%
\[
\bar{\psi}\phi_{2}\psi:%
\begin{array}
[c]{c}%
\begin{picture}(62,24) (59,-19)
\SetWidth{1.0}
\SetColor{Black}
\Line
[arrow,arrowpos=0.5,arrowlength=3,arrowwidth=1.2,arrowinset=0.2](120,-16)(90,-16)
\Line
[arrow,arrowpos=0.5,arrowlength=3,arrowwidth=1.2,arrowinset=0.2](90,-16)(60,-16)
\Text(100,-6)[]{\normalsize{\Black{$\phi_2$}}}
\SetWidth{0.5}
\Line[dash,dashsize=0.3](90,-16)(90,4)
\end{picture}%
\end{array}
=f\left(  R-L\right)  =f\gamma_{5}%
\]%
\[
\bar{\psi}H\psi:%
\begin{array}
[c]{c}%
\begin{picture}(62,24) (59,-19)
\SetWidth{1.0}
\SetColor{Black}
\Line
[arrow,arrowpos=0.5,arrowlength=3,arrowwidth=1.2,arrowinset=0.2](120,-16)(90,-16)
\Line
[arrow,arrowpos=0.5,arrowlength=3,arrowwidth=1.2,arrowinset=0.2](90,-16)(60,-16)
\Text(100,-6)[]{\normalsize{\Black{$H$}}}
\SetWidth{0.5}
\Line[dash,dashsize=0.3](90,-16)(90,4)
\end{picture}%
\end{array}
=-if
\]%
\[
HA^{\mu}\phi_{2}:%
\begin{array}
[c]{c}%
\begin{picture}(122,66) (25,0)
\SetWidth{0.5}
\SetColor{Black}
\Photon(90,46)(90,26){2.5}{4}
\Text(100,36)[]{\normalsize{\Black{$\mu$}}}
\Line
[dash,dashsize=0.3,arrow,arrowpos=0.5,arrowlength=2.5,arrowwidth=1,arrowinset=0.2](90,26)(60,16)
\Line
[dash,dashsize=0.3,arrow,arrowpos=0.5,arrowlength=2.5,arrowwidth=1,arrowinset=0.2](120,16)(90,26)
\Text(70,6)[]{\normalsize{\Black{$H$}}}
\Text(110,6)[]{\normalsize{\Black{$\phi_2$}}}
\Text(130,16)[]{\normalsize{\Black{$p$}}}
\Text(90,56)[]{\normalsize{\Black{$k$}}}
\Text(40,16)[]{\normalsize{\Black{$p+k$}}}
\end{picture}%
\end{array}
=g\left(  2p^{\mu}+k^{\mu}\right)
\]

\subsection{1-Loop Counter Terms}

\begin{figure}
[tbh]
\begin{center}
\includegraphics[
height=0.8579in,
width=5.6299in
]%
{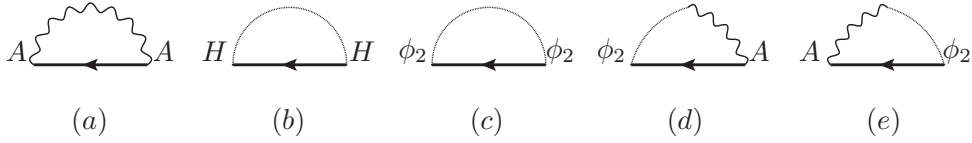}%
\caption{Fermion Self-Energy Diagrams}%
\label{bm-o2}%
\end{center}
\end{figure}

\subsubsection{Figure \ref{bm-o2}: $\bar{\psi}\psi$ Self-Energy Diagrams}

The possible diagrams that may contribute to the fermion self-energy are
depicted in Figure \ref{bm-o2}.%

In each diagram of Figure $\ref{bm-o2}$, the horizontal line signifies an
internal fermion line and the wavy line is a vector meson line.%

\noindent
Diagram $\left(  a\right)  $

This diagram has been discussed thoroughly in Sec. \ref{exsc}. The Feynman
amplitude in the BM scheme is denoted by $\Gamma_{\ref{bm-o2}\left(  a\right)
}^{BM}$.%
\[
\Gamma_{\ref{bm-o2}\left(  a\right)  }^{BM}=-ig^{2}\int\frac{d^{n}\ell
}{\left(  2\pi\right)  ^{n}}D\left(  A^{\mu},A^{\nu};\ell\right)  R\gamma
^{\mu}L\frac{1}{\not \ell +\not p  -m}R\gamma^{\nu}L
\]
The corresponding amplitude in the rightmost $\gamma_{5}$ scheme is denoted by
$\Gamma_{\ref{bm-o2}\left(  a\right)  }^{R5}$.
\[
\Gamma_{\ref{bm-o2}\left(  a\right)  }^{R5}=-ig^{2}\int\frac{d^{n}\ell
}{\left(  2\pi\right)  ^{n}}D\left(  A^{\mu},A^{\nu};\ell\right)  \gamma^{\mu
}\frac{\not \ell +\not p  }{\left(  \ell+p\right)  ^{2}-m^{2}}\gamma^{\nu}L
\]
The difference has been shown in Sec. \ref{exsc} to be%
\[
\lim_{n\rightarrow4}\left(  \Gamma_{\ref{bm-o2}\left(  a\right)  }^{R5}%
-\Gamma_{\ref{bm-o2}\left(  a\right)  }^{BM}\right)  =-\frac{1}{\left(
4\pi\right)  ^{2}}i\frac{g^{2}}{3}\left(  1+2\alpha\right)  \not p  L
\]

\noindent
Diagram $\left(  b\right)  $

There is no $\gamma_{5}$ in the BM amplitude $\Gamma_{\ref{bm-o2}\left(
b\right)  }^{BM}$. The rightmost $\gamma_{5}$ amplitude $\Gamma_{\ref{bm-o2}%
\left(  b\right)  }^{R5}$ is the same as $\Gamma_{\ref{bm-o2}\left(  b\right)
}^{BM}$ and no finite counter term is generated.%
\[
\Gamma_{\ref{bm-o2}\left(  b\right)  }^{R5}-\Gamma_{\ref{bm-o2}\left(
b\right)  }^{BM}=0
\]

\noindent
Diagram $\left(  c\right)  $%
\[
\Gamma_{\ref{bm-o2}\left(  c\right)  }^{BM}=f^{2}\int\frac{d^{n}\ell}{\left(
2\pi\right)  ^{n}}D\left(  \phi_{2},\phi_{2},\ell\right)  \gamma_{5}\frac
{i}{\not \ell +\not p  -m}\gamma_{5}%
\]%
\[
\Gamma_{\ref{bm-o2}\left(  c\right)  }^{R5}=f^{2}\int\frac{d^{n}\ell}{\left(
2\pi\right)  ^{n}}D\left(  \phi_{2},\phi_{2},\ell\right)  \frac{i}%
{-\not \ell -\not p  -m}%
\]%
\begin{align*}
&  \lim_{n\rightarrow4}\left(  \Gamma_{\ref{bm-o2}\left(  c\right)  }%
^{R5}-\Gamma_{\ref{bm-o2}\left(  c\right)  }^{BM}\right) \\
&  =if^{2}\int\frac{d^{n}\ell}{\left(  2\pi\right)  ^{n}}D\left(  \phi
_{2},\phi_{2},\ell\right)  \left(  \frac{-\not \ell -\not p  +m-\gamma
_{5}\left(  \not \ell +\not p  +m\right)  \gamma_{5}}{\left(  \ell+p\right)
^{2}-m^{2}}\right) \\
&  =if^{2}\int\frac{d^{n}\ell}{\left(  2\pi\right)  ^{n}}D\left(  \phi
_{2},\phi_{2},\ell\right)  \frac{-2\not \ell _{\Delta}}{\left(  \ell+p\right)
^{2}-m^{2}}=0
\end{align*}

\noindent
Diagram $\left(  d\right)  $%
\[
\Gamma_{\ref{bm-o2}\left(  d\right)  }^{BM}=\left(  -ig\right)  f\int
\frac{d^{n}\ell}{\left(  2\pi\right)  ^{n}}D\left(  A^{\mu},\phi_{2}%
;\ell\right)  \gamma_{5}\frac{i}{\not \ell +\not p  -m}R\gamma^{\mu}L
\]%
\[
\Gamma_{\ref{bm-o2}\left(  d\right)  }^{R5}=\left(  -ig\right)  f\int
\frac{d^{n}\ell}{\left(  2\pi\right)  ^{n}}D\left(  A^{\mu},\phi_{2}%
;\ell\right)  \frac{i}{-\not \ell -\not p  -m}\gamma^{\mu}L
\]%
\begin{align*}
&  \lim_{n\rightarrow4}\left(  \Gamma_{\ref{bm-o2}\left(  d\right)  }%
^{R5}-\Gamma_{\ref{bm-o2}\left(  d\right)  }^{BM}\right) \\
&  =gf\lim_{n\rightarrow4}\int\frac{d^{n}\ell}{\left(  2\pi\right)  ^{n}%
}D\left(  A^{\mu},\phi_{2};\ell\right)  \frac{-\not \ell -\gamma_{5}%
\not \ell R}{\ell^{2}-m^{2}}\gamma^{\mu}L\\
&  =\alpha\left(  M-\Lambda\right)  gf\lim_{n\rightarrow4}\int\frac{d^{n}\ell
}{\left(  2\pi\right)  ^{n}}\frac{-\not \ell -\gamma_{5}\not \ell R}{\left(
\ell^{2}-m\right)  ^{3}}\not \ell L\\
&  =-\alpha\left(  M-\Lambda\right)  gf\lim_{n\rightarrow4}\int\frac{d^{n}%
\ell}{\left(  2\pi\right)  ^{n}}\frac{\ell_{\Delta}^{2}}{\left(  \ell
^{2}-m\right)  ^{3}}L=\frac{1}{\left(  4\pi\right)  ^{2}}\frac{i}{2}%
\alpha\left(  M-\Lambda\right)  fgL
\end{align*}

\noindent
Diagram $\left(  e\right)  $%
\[
\Gamma_{\ref{bm-o2}\left(  e\right)  }^{BM}=\left(  -ig\right)  f\int
\frac{d^{n}\ell}{\left(  2\pi\right)  ^{n}}D\left(  \phi_{2},A^{\mu}%
;\ell\right)  R\gamma^{\mu}L\frac{i}{\not \ell +\not p  -m}\gamma_{5}%
\]%
\[
\Gamma_{\ref{bm-o2}\left(  e\right)  }^{R5}=gf\int\frac{d^{n}\ell}{\left(
2\pi\right)  ^{n}}D\left(  \phi_{2},A^{\mu};\ell\right)  \gamma^{\mu}%
\frac{\left(  \not \ell +\not p  \right)  R-mL}{\left(  \ell+p\right)
^{2}-m^{2}}%
\]%
\begin{align*}
&  \lim_{n\rightarrow4}\left(  \Gamma_{\ref{bm-o2}\left(  e\right)  }%
^{R5}-\Gamma_{\ref{bm-o2}\left(  e\right)  }^{BM}\right) \\
&  =gf\int\frac{d^{n}\ell}{\left(  2\pi\right)  ^{n}}D\left(  \phi_{2},A^{\mu
};\ell\right)  \frac{\gamma^{\mu}\not \ell R-R\gamma^{\mu}L\not \ell
\gamma_{5}}{\ell^{2}-m^{2}}\\
&  =-\alpha\left(  M-\Lambda\right)  gf\int\frac{d^{n}\ell}{\left(
2\pi\right)  ^{n}}\frac{\ell_{\Delta}^{2}R}{\left(  \ell^{2}-m\right)  ^{3}%
}=\frac{1}{\left(  4\pi\right)  ^{2}}\frac{i}{2}\alpha\left(  M-\Lambda
\right)  fgR
\end{align*}%
\noindent
Summary

The amplitudes and finite counter terms due to diagrams in Figure
$\ref{bm-o2}$ are tabulated in Table \ref{ct-o2}.

$\ $%
\begin{table}[htp] \centering
\begin{tabular}
[c]{|l|l|l|}\hline
Figure & $\left(  4\pi\right)  ^{2}\times\left(  \Gamma^{R5}-\Gamma
^{BM}\right)  $ & $\left(  4\pi\right)  ^{2}\times$ Counter Term\\\hline
$\ref{bm-o2}\left(  a\right)  $ & $-i\frac{g^{2}}{3}\left(  1+2\alpha\right)
\not p  L$ & $-\frac{g^{2}}{3}\left(  1+2\alpha\right)  \bar{\psi
}Ri\not \partial L\psi$\\
$\ref{bm-o2}\left(  b\right)  $ & $0$ & $0$\\
$\ref{bm-o2}\left(  c\right)  $ & $0$ & $0$\\
$\ref{bm-o2}\left(  d\right)  $ & $\frac{i}{2}\alpha\left(  M-\Lambda\right)
fgL$ & $\frac{1}{2}\alpha\left(  M-\Lambda\right)  fg\bar{\psi}L\psi$\\
$\ref{bm-o2}\left(  e\right)  $ & $\frac{i}{2}\alpha\left(  M-\Lambda\right)
fgR$ & $\frac{1}{2}\alpha\left(  M-\Lambda\right)  fg\bar{\psi}R\psi$\\\hline
\end{tabular}
\caption{Counter terms due to diagrams in Figure \ref{bm-o2}\label{ct-o2}}\label{tb1}%
\end{table}%

\begin{figure}
[tbh]
\begin{center}
\includegraphics[
height=1.0585in,
width=5.2788in
]%
{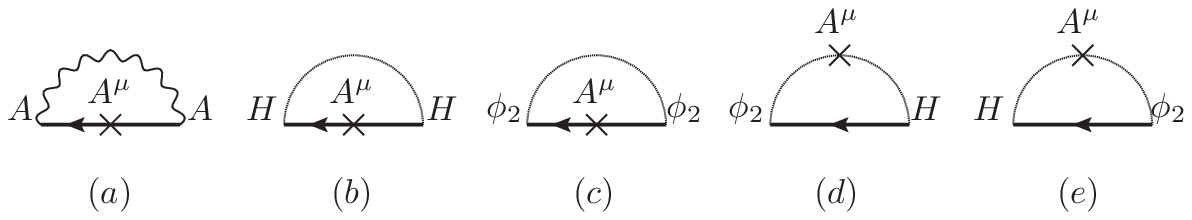}%
\caption{Diagrams for $\bar{\psi}A\psi$ Vertex}%
\label{bm-o3A}%
\end{center}
\end{figure}

\subsubsection{Figure $\ref{bm-o3A}$: $\bar{\psi}A\psi$ Vertex Diagrams}%

In Figure $\ref{bm-o3A}$, $\mu$ $\in\left\{  0,1,2,3\right\}  $ is a
polarization in first 4 dimensions.%

\noindent
Diagram $\left(  a\right)  $%
\[
\Gamma_{\ref{bm-o3A}\left(  a\right)  }^{BM}=\left(  -ig\right)  ^{3}\int
\frac{d^{n}\ell}{\left(  2\pi\right)  ^{n}}D\left(  A_{\rho},A_{\sigma}%
,\ell\right)  R\gamma^{\rho}L\frac{i}{\not \ell +\not p  +\not k  -m}%
R\gamma^{\mu}L\frac{i}{\not \ell +\not p  -m}R\gamma^{\sigma}L
\]%
\[
\Gamma_{\ref{bm-o3A}\left(  a\right)  }^{R5}=-ig^{3}\int\frac{d^{n}\ell
}{\left(  2\pi\right)  ^{n}}D\left(  A_{\rho},A_{\sigma},\ell\right)
\gamma^{\rho}\frac{\not \ell +\not p  +\not k  }{\left(  \ell+p+k\right)
^{2}-m^{2}}\gamma^{\mu}\frac{\not \ell +\not p  }{\left(  \ell+p\right)
^{2}-m^{2}}\gamma^{\sigma}L
\]%
\begin{align*}
&  \Gamma_{\ref{bm-o3A}\left(  a\right)  }^{R5}-\Gamma_{\ref{bm-o3A}\left(
a\right)  }^{BM}\\
&  =-ig^{3}\int\frac{d^{n}\ell}{\left(  2\pi\right)  ^{n}}D\left(  A_{\rho
},A_{\sigma},\ell\right)  \frac{\gamma^{\rho}\not \ell \gamma^{\mu}%
\not \ell \gamma^{\sigma}L-R\gamma^{\rho}L\not \ell R\gamma^{\mu}%
L\not \ell R\gamma^{\sigma}L}{\left(  \ell^{2}-m^{2}\right)  ^{2}}\\
&  =-g^{3}\int\frac{d^{n}\ell}{\left(  2\pi\right)  ^{n}}\frac{\left(
g^{\rho\sigma}+\left(  \alpha-1\right)  \frac{\ell^{\rho}\ell^{\sigma}}%
{\ell^{2}-m^{2}}\right)  \left(  \gamma^{\rho}\not \ell \gamma^{\mu}%
\not \ell \gamma^{\sigma}-\underline{\gamma}^{\rho}\underline{\not \ell
}\gamma^{\mu}\underline{\not \ell }\underline{\gamma}^{\sigma}\right)
L}{\left(  \ell^{2}-m^{2}\right)  ^{4}}\\
&  =-g^{3}\int\frac{d^{n}\ell}{\left(  2\pi\right)  ^{n}}\frac{\left(
\frac{\left(  2-n\right)  ^{2}}{n}\ell^{2}-\underline{\ell}^{2}+\left(
\alpha-1\right)  \frac{\left(  \ell^{4}-\underline{\ell}^{4}\right)  }%
{\ell^{2}-m^{2}}\right)  \gamma^{\mu}L}{\left(  \ell^{2}-m^{2}\right)  ^{4}}\\
&  =\frac{1}{\left(  4\pi\right)  ^{2}}\frac{i}{6}g^{3}\left(  7+5\alpha
\right)  \gamma^{\mu}L
\end{align*}

\noindent
Diagram $\left(  b\right)  $%
\[
\Gamma_{\ref{bm-o3A}\left(  b\right)  }^{BM}=\left(  -ig\right)  \left(
-if\right)  ^{2}\int\frac{d^{n}\ell}{\left(  2\pi\right)  ^{n}}D\left(
H,H;\ell\right)  \frac{i}{\not \ell +\not p  +\not k  -m}R\gamma^{\mu}%
L\frac{i}{\not \ell +\not p  -m}%
\]%
\[
\Gamma_{\ref{bm-o3A}\left(  b\right)  }^{R5}=-igf^{2}\int\frac{d^{n}\ell
}{\left(  2\pi\right)  ^{n}}D\left(  H,H;\ell\right)  \frac{1}{\not \ell
+\not p  +\not k  -m}\gamma^{\mu}\frac{\left(  \not \ell +\not p  \right)
R+mL}{\left(  \ell+p\right)  ^{2}-m^{2}}%
\]%
\begin{align*}
&  \Gamma_{\ref{bm-o3A}\left(  b\right)  }^{R5}-\Gamma_{\ref{bm-o3A}\left(
b\right)  }^{BM}\\
&  =gf^{2}\int\frac{d^{n}\ell}{\left(  2\pi\right)  ^{n}}\frac{1}{\ell
^{2}-m^{2}}\frac{1}{\not \ell -m}\left(  \frac{\gamma^{\mu}\not \ell R}%
{\ell^{2}-m^{2}}-\frac{R\gamma^{\mu}L\not \ell }{\ell^{2}-m^{2}}\right) \\
&  =-gf^{2}\int\frac{d^{n}\ell}{\left(  2\pi\right)  ^{n}}\frac{\ell_{\Delta
}^{2}\gamma_{5}}{\left(  \ell^{2}-m^{2}\right)  ^{3}}=\frac{1}{\left(
4\pi\right)  ^{2}}\frac{1}{2}igf^{2}\gamma^{\mu}\gamma_{5}%
\end{align*}

\noindent
Diagram $\left(  c\right)  $%
\[
\Gamma_{\ref{bm-o3A}\left(  c\right)  }^{BM}=-igf^{2}\int\frac{d^{n}\ell
}{\left(  2\pi\right)  ^{n}}D\left(  \phi_{2},\phi_{2};\ell\right)  \gamma
_{5}\frac{i}{\not \ell +\not p  +\not k  -m}R\gamma^{\mu}L\frac{i}%
{\not \ell +\not p  -m}\gamma_{5}%
\]%
\[
\Gamma_{\ref{bm-o3A}\left(  c\right)  }^{R5}=-igf^{2}\int\frac{d^{n}\ell
}{\left(  2\pi\right)  ^{n}}D\left(  \phi_{2},\phi_{2};\ell\right)  \frac
{1}{\not \ell +\not p  +\not k  +m}\gamma^{\mu}\frac{\left(  \not \ell
+\not p  \right)  R-mL}{\left(  \ell+p\right)  ^{2}-m^{2}}%
\]%
\begin{align*}
&  \lim_{n\rightarrow4}\left(  \Gamma_{\ref{bm-o3A}\left(  c\right)  }%
^{R5}-\Gamma_{\ref{bm-o3A}\left(  c\right)  }^{BM}\right) \\
&  =gf^{2}\int\frac{d^{n}\ell}{\left(  2\pi\right)  ^{n}}\frac{1}{\ell
^{2}-m^{2}}\left(  \frac{1}{\not \ell +m}\gamma^{\mu}\frac{\not \ell R}%
{\ell^{2}-m^{2}}+\gamma_{5}\frac{1}{\not \ell -m}R\gamma^{\mu}L\frac
{1}{\not \ell -m}\gamma_{5}\right) \\
&  =-gf^{2}\int\frac{d^{n}\ell}{\left(  2\pi\right)  ^{n}}\frac{\ell_{\Delta
}^{2}\gamma_{5}}{\left(  \ell^{2}-m^{2}\right)  ^{3}}=\frac{1}{\left(
4\pi\right)  ^{2}}i\frac{1}{2}gf^{2}\gamma^{\mu}\gamma_{5}%
\end{align*}

\noindent
Diagram $\left(  d\right)  $%
\[
\Gamma_{\ref{bm-o3A}\left(  d\right)  }^{BM}=gf^{2}\int\frac{d^{n}\ell
}{\left(  2\pi\right)  ^{n}}D\left(  \phi_{2},\phi_{2};\ell-k\right)  \left(
2\ell-k\right)  ^{\mu}D\left(  H,H;\ell\right)  \gamma_{5}\frac{1}%
{\not \ell +\not p  -m}%
\]%
\[
\Gamma_{\ref{bm-o3A}\left(  d\right)  }^{R5}=gf^{2}\int\frac{d^{n}\ell
}{\left(  2\pi\right)  ^{n}}D\left(  \phi_{2},\phi_{2};\ell-k\right)  \left(
2\ell-k\right)  ^{\mu}D\left(  H,H;\ell\right)  \frac{-1}{\not \ell +\not p
+m}\gamma_{5}%
\]%
\begin{align*}
&  \lim_{n\rightarrow4}\left(  \Gamma_{\ref{bm-o3A}\left(  d\right)  }%
^{R5}-\Gamma_{\ref{bm-o3A}\left(  d\right)  }^{BM}\right) \\
&  =gf^{2}\int\frac{d^{n}\ell}{\left(  2\pi\right)  ^{n}}D\left(  \phi
_{2},\phi_{2};\ell\right)  D\left(  H,H;\ell\right)  2\ell^{\mu}\left(
\frac{-1}{\not \ell +m}\gamma_{5}-\gamma_{5}\frac{1}{\not \ell -m}\right) \\
&  =-4gf^{2}\int\frac{d^{n}\ell}{\left(  2\pi\right)  ^{n}}D\left(  \phi
_{2},\phi_{2};\ell\right)  D\left(  H,H;\ell\right)  \frac{\ell^{\mu
}\not \ell _{\Delta}}{\left(  \ell^{2}-m^{2}\right)  }=0
\end{align*}

\noindent
Diagram $\left(  e\right)  $%
\[
\Gamma_{\ref{bm-o3A}\left(  e\right)  }^{BM}=gf^{2}\int\frac{d^{n}\ell
}{\left(  2\pi\right)  ^{n}}D\left(  H,H;\ell-k\right)  \left(  k-2\ell
\right)  ^{\mu}D\left(  \phi_{2},\phi_{2};\ell\right)  \frac{1}{\not \ell
+\not p  -m}\gamma_{5}%
\]%
\[
\Gamma_{\ref{bm-o3A}\left(  e\right)  }^{R5}=gf^{2}\int\frac{d^{n}\ell
}{\left(  2\pi\right)  ^{n}}D\left(  H,H;\ell-k\right)  \left(  k-2\ell
\right)  ^{\mu}D\left(  \phi_{2},\phi_{2};\ell\right)  \frac{1}{\not \ell
+\not p  -m}\gamma_{5}%
\]%
\[
\Gamma_{\ref{bm-o3A}\left(  e\right)  }^{R5}-\Gamma_{\ref{bm-o3A}\left(
e\right)  }^{BM}=0
\]%
\noindent
Summary

The amplitudes and finite counter terms due to diagrams in Figure
$\ref{bm-o3A}$ are tabulated in Table \ref{ct-o3A}.

$\ $%
\begin{table}[htp] \centering
\begin{tabular}
[c]{|l|l|l|}\hline
Figure & $\left(  4\pi\right)  ^{2}\times\left(  \Gamma^{R5}-\Gamma
^{BM}\right)  $ & $\left(  4\pi\right)  ^{2}\times$ Counter Term\\\hline
$\ref{bm-o3A}\left(  a\right)  $ & $\frac{i}{6}g^{3}\left(  7+5\alpha\right)
\gamma^{\mu}L$ & $\frac{1}{6}g^{3}\left(  7+5\alpha\right)  \bar{\psi}R\not A
L\psi$\\
$\ref{bm-o3A}\left(  b\right)  $ & $\frac{1}{2}igf^{2}\gamma^{\mu}\gamma_{5}$
& $\frac{1}{2}gf^{2}\bar{\psi}\not A  \gamma_{5}\psi$\\
$\ref{bm-o3A}\left(  c\right)  $ & $\frac{1}{2}igf^{2}\gamma^{\mu}\gamma_{5}$
& $\frac{1}{2}gf^{2}\bar{\psi}\not A  \gamma_{5}\psi$\\
$\ref{bm-o3A}\left(  d\right)  $ & $0$ & $0$\\
$\ref{bm-o3A}\left(  e\right)  $ & $0$ & $0$\\\hline
\end{tabular}
\caption{Counter terms due to diagrams in Figure \ref{bm-o3A}\label{ct-o3A}}\label{tb2}%
\end{table}%

\begin{figure}
[tbh]
\begin{center}
\includegraphics[
height=1.0585in,
width=5.2788in
]%
{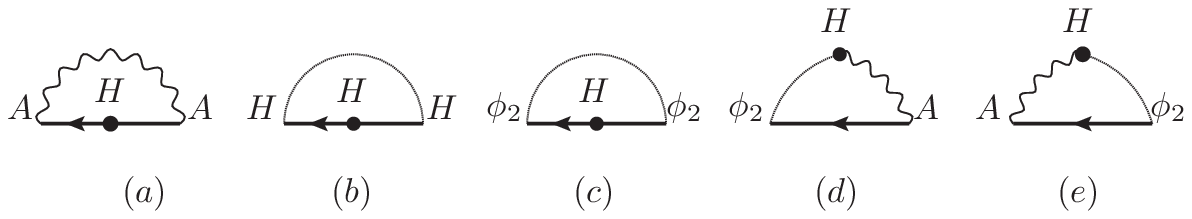}%
\caption{Diagrams for $\bar{\psi}H\psi$ Vertex}%
\label{bm-o3H}%
\end{center}
\end{figure}

\subsubsection{Figure $\ref{bm-o3H}$: $\bar{\psi}H\psi$ Vertex Diagrams}%

\noindent
Diagram $\left(  a\right)  $%
\[
\Gamma_{\ref{bm-o3H}\left(  a\right)  }^{BM}=-ifg^{2}\int\frac{d^{n}\ell
}{\left(  2\pi\right)  ^{n}}D\left(  A_{\mu},A_{\nu};\ell\right)  R\gamma
^{\mu}L\frac{1}{\not \ell +\not p  +\not k  -m}\frac{1}{\not \ell +\not p
-m}R\gamma^{\nu}L
\]%
\[
\Gamma_{\ref{bm-o3H}\left(  a\right)  }^{R5}=-ifg^{2}\int\frac{d^{n}\ell
}{\left(  2\pi\right)  ^{n}}D\left(  A_{\mu},A_{\nu};\ell\right)  \gamma^{\mu
}\frac{m\left(  2\not \ell +2\not p  +\not k  \right)  }{\left(  \left(
\ell+p+k\right)  ^{2}-m^{2}\right)  \left(  \left(  \ell+p\right)  ^{2}%
-m^{2}\right)  }\gamma^{\nu}L
\]%
\begin{align*}
&  \lim_{n\rightarrow4}\left(  \Gamma_{\ref{bm-o3H}\left(  a\right)  }%
^{R5}-\Gamma_{\ref{bm-o3H}\left(  a\right)  }^{BM}\right) \\
&  =2fg^{2}\int\frac{d^{n}\ell}{\left(  2\pi\right)  ^{n}}\frac{g_{\mu\nu
}+\left(  \alpha-1\right)  \frac{\ell_{\mu}\ell_{\nu}}{\ell^{2}}}{\left(
\ell^{2}-m^{2}\right)  ^{3}}\left(
\begin{array}
[c]{c}%
m\gamma^{\mu}\not \ell \gamma^{\nu}\\
-\underline{\gamma}^{\mu}\left(  m+\not \ell _{\Delta}\right)  \underline
{\not \ell }\underline{\gamma}^{\nu}%
\end{array}
\right)  L\\
&  =-2fg^{2}\int\frac{d^{n}\ell}{\left(  2\pi\right)  ^{n}}\frac{g_{\mu\nu
}+\left(  \alpha-1\right)  \frac{\ell_{\mu}\ell_{\nu}}{\ell^{2}}}{\left(
\ell^{2}-m^{2}\right)  ^{3}}\underline{\gamma}^{\mu}\not \ell _{\Delta
}\underline{\not \ell }\underline{\gamma}^{\nu}L=0
\end{align*}

\noindent
Diagram $\left(  b\right)  $

No $\gamma_{5}$ is involved.%
\[
\Gamma_{\ref{bm-o3H}\left(  b\right)  }^{R5}-\Gamma_{\ref{bm-o3H}\left(
b\right)  }^{BM}=0
\]

\noindent
Diagram $\left(  c\right)  $%
\[
\Gamma_{\ref{bm-o3H}\left(  c\right)  }^{BM}=-if^{3}\int\frac{d^{n}\ell
}{\left(  2\pi\right)  ^{n}}D\left(  \phi_{2},\phi_{2};\ell\right)  \gamma
_{5}\frac{1}{\not \ell +\not p  +\not k  -m}\frac{1}{\not \ell +\not p
-m}\gamma_{5}%
\]%
\[
\Gamma_{\ref{bm-o3H}\left(  c\right)  }^{R5}=-if^{3}\int\frac{d^{n}\ell
}{\left(  2\pi\right)  ^{n}}D\left(  \phi_{2},\phi_{2};\ell\right)  \frac
{1}{\not \ell +\not p  +\not k  +m}\frac{1}{\not \ell +\not p  +m}%
\]%
\begin{align*}
&  \lim_{n\rightarrow4}\left(  \Gamma_{\ref{bm-o3H}\left(  c\right)  }%
^{R5}-\Gamma_{\ref{bm-o3H}\left(  c\right)  }^{BM}\right) \\
&  =f^{3}\int\frac{d^{n}\ell}{\left(  2\pi\right)  ^{n}}\frac{1}{\left(
\ell^{2}-m^{2}\right)  }\left(  \left(  \frac{1}{\not \ell +m}\right)
^{2}-\gamma_{5}\left(  \frac{1}{\not \ell -m}\right)  ^{2}\gamma_{5}\right) \\
&  =f^{3}\int\frac{d^{n}\ell}{\left(  2\pi\right)  ^{n}}\frac{1}{\left(
\ell^{2}-m^{2}\right)  ^{3}}\left(  \ell^{2}\gamma_{5}-\gamma_{5}\ell
^{2}\gamma_{5}\right)  =0
\end{align*}

\noindent
Diagram $\left(  d\right)  $%
\[
\Gamma_{\ref{bm-o3H}\left(  d\right)  }^{BM}=g^{2}f\int\frac{d^{n}\ell
}{\left(  2\pi\right)  ^{n}}D\left(  \phi_{2},\phi_{2};\ell\right)  D\left(
A_{\mu},A_{\nu};\ell+k\right)  \left(  \ell-k\right)  ^{\mu}\gamma_{5}\frac
{1}{\not \ell +\not p  +\not k  -m}R\gamma^{\nu}L
\]

\[
\Gamma_{\ref{bm-o3H}\left(  d\right)  }^{R5}=-g^{2}f\int\frac{d^{n}\ell
}{\left(  2\pi\right)  ^{n}}D\left(  \phi_{2},\phi_{2};\ell\right)  D\left(
A_{\mu},A_{\nu};\ell+k\right)  \left(  \ell-k\right)  ^{\mu}\frac
{1}{\not \ell +\not p  +\not k  +m}\gamma^{\nu}L
\]%
\begin{align*}
&  \lim_{n\rightarrow4}\left(  \Gamma_{\ref{bm-o3H}\left(  d\right)  }%
^{R5}-\Gamma_{\ref{bm-o3H}\left(  d\right)  }^{BM}\right) \\
&  =-g^{2}f\int\frac{d^{n}\ell}{\left(  2\pi\right)  ^{n}}D\left(  \phi
_{2},\phi_{2};\ell\right)  D\left(  A_{\mu},A_{\nu};\ell\right)  \ell^{\mu
}\left(  \frac{1}{\not \ell +m}\gamma^{\nu}L+\gamma_{5}\frac{1}{\not \ell
-m}R\gamma^{\nu}L\right) \\
&  =-g^{2}f\int\frac{d^{n}\ell}{\left(  2\pi\right)  ^{n}}\frac{\alpha
}{\left(  \ell^{2}-m^{2}\right)  ^{3}}\left(  \ell^{2}L+\gamma_{5}%
\not \ell R\not \ell L\right) \\
&  =-g^{2}f\int\frac{d^{n}\ell}{\left(  2\pi\right)  ^{n}}\frac{\alpha
\ell_{\Delta}^{2}}{\left(  \ell^{2}-m^{2}\right)  ^{3}}L=\frac{1}{\left(
4\pi\right)  ^{2}}i\frac{\alpha g^{2}f}{2}L
\end{align*}

\noindent
Diagram $\left(  e\right)  $%
\[
\Gamma_{\ref{bm-o3H}\left(  e\right)  }^{BM}=gf\int\frac{d^{n}\ell}{\left(
2\pi\right)  ^{n}}D\left(  A_{\mu},A_{\nu};\ell\right)  \left(  -g\left(
\ell+2k\right)  ^{\mu}\right)  D\left(  \phi_{2},\phi_{2};\ell+k\right)
R\gamma^{\nu}L\frac{1}{\not \ell +\not p  +\not k  -m}\gamma_{5}%
\]%
\[
\Gamma_{\ref{bm-o3H}\left(  e\right)  }^{R5}=gf\int\frac{d^{n}\ell}{\left(
2\pi\right)  ^{n}}D\left(  A_{\mu},A_{\nu};\ell\right)  \left(  -g\left(
\ell+2k\right)  ^{\mu}\right)  D\left(  \phi_{2},\phi_{2};\ell+k\right)
\gamma^{\nu}\frac{\left(  \not \ell +\not p  +\not k  \right)  R-mL}{\left(
\ell+p+k\right)  ^{2}-m^{2}}%
\]%
\begin{align*}
&  \lim_{n\rightarrow4}\left(  \Gamma_{\ref{bm-o3H}\left(  e\right)  }%
^{R5}-\Gamma_{\ref{bm-o3H}\left(  e\right)  }^{BM}\right) \\
&  =-g^{2}f\int\frac{d^{n}\ell}{\left(  2\pi\right)  ^{n}}\frac{\alpha
\ell_{\Delta}^{2}}{\left(  \ell^{2}-m^{2}\right)  ^{3}}\left(  \ell
^{2}R-R\not \ell L\not \ell \gamma_{5}\right) \\
&  =-g^{2}f\int\frac{d^{n}\ell}{\left(  2\pi\right)  ^{n}}\frac{\alpha
\ell_{\Delta}^{2}}{\left(  \ell^{2}-m^{2}\right)  ^{3}}R=\frac{1}{\left(
4\pi\right)  ^{2}}i\frac{\alpha g^{2}f}{2}R
\end{align*}%
\noindent
Summary

The amplitudes and finite counter terms due to diagrams in Figure
$\ref{bm-o3H}$ are tabulated in Table \ref{ct-o3H}.

$\ $%
\begin{table}[htp] \centering
\begin{tabular}
[c]{|l|l|l|}\hline
Figure & $\left(  4\pi\right)  ^{2}\times\left(  \Gamma^{R5}-\Gamma
^{BM}\right)  $ & $\left(  4\pi\right)  ^{2}\times$ Counter Term\\\hline
$\ref{bm-o3H}\left(  a\right)  $ & $0$ & $0$\\
$\ref{bm-o3H}\left(  b\right)  $ & $0$ & $0$\\
$\ref{bm-o3H}\left(  c\right)  $ & $0$ & $0$\\
$\ref{bm-o3H}\left(  d\right)  $ & $i\frac{\alpha g^{2}f}{2}L$ & $\frac{\alpha
g^{2}f}{2}\bar{\psi}HL\psi$\\
$\ref{bm-o3H}\left(  e\right)  $ & $i\frac{\alpha g^{2}f}{2}R$ & $\frac{\alpha
g^{2}f}{2}\bar{\psi}HR\psi$\\\hline
\end{tabular}
\caption{Counter terms due to diagrams in Figure \ref{bm-o3H}\label{ct-o3H}}\label{tb3}%
\end{table}%

\begin{figure}
[tbh]
\begin{center}
\includegraphics[
height=1.0585in,
width=5.2788in
]%
{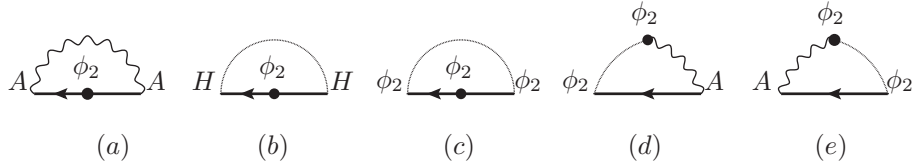}%
\caption{Diagrams for $\bar{\psi}\phi_{2}\psi$ Vertex}%
\label{bm-o3phi2}%
\end{center}
\end{figure}

\subsubsection{Figure $\ref{bm-o3phi2}$: $\bar{\psi}\phi_{2}\psi$ Vertex
Diagrams}%

\noindent
Diagram $\left(  a\right)  $%
\[
\Gamma_{\ref{bm-o3phi2}\left(  a\right)  }^{BM}=fg^{2}\int\frac{d^{n}\ell
}{\left(  2\pi\right)  ^{n}}D\left(  A_{\mu},A_{\nu};\ell\right)  R\gamma
^{\mu}L\frac{1}{\not \ell +\not p  +\not k  -m}\gamma_{5}\frac{1}%
{\not \ell +\not p  -m}R\gamma^{\nu}L
\]%
\[
\Gamma_{\ref{bm-o3phi2}\left(  a\right)  }^{R5}=fg^{2}\int\frac{d^{n}\ell
}{\left(  2\pi\right)  ^{n}}D\left(  A_{\mu},A_{\nu};\ell\right)  \gamma^{\mu
}\frac{m\not k  }{\left(  \left(  \ell+p+k\right)  ^{2}-m^{2}\right)  \left(
\left(  \ell+p\right)  ^{2}-m^{2}\right)  }\gamma^{\nu}L
\]%
\begin{align*}
&  \lim_{n\rightarrow4}\left(  \Gamma_{\ref{bm-o3phi2}\left(  a\right)  }%
^{R5}-\Gamma_{\ref{bm-o3phi2}\left(  a\right)  }^{BM}\right) \\
&  =-ifg^{2}\int\frac{d^{n}\ell}{\left(  2\pi\right)  ^{n}}\frac{g_{\mu\nu
}+\left(  \alpha-1\right)  \frac{\ell_{\mu}\ell_{\nu}}{\ell^{2}}}{\left(
\ell^{2}-m^{2}\right)  ^{3}}\left(  -2m\underline{\gamma}^{\mu}\not \ell
_{\Delta}\underline{\gamma}^{\nu}\right)  L=0
\end{align*}

\noindent
Diagram $\left(  b\right)  $%
\[
\Gamma_{\ref{bm-o3phi2}\left(  b\right)  }^{BM}=f^{3}\int\frac{d^{n}\ell
}{\left(  2\pi\right)  ^{n}}D\left(  \phi_{2},\phi_{2};\ell\right)  \frac
{1}{\not \ell +\not p  +\not k  -m}\gamma_{5}\frac{1}{\not \ell +\not p  -m}%
\]%
\[
\Gamma_{\ref{bm-o3phi2}\left(  b\right)  }^{R5}=-f^{3}\int\frac{d^{n}\ell
}{\left(  2\pi\right)  ^{n}}D\left(  \phi_{2},\phi_{2};\ell\right)  \frac
{1}{\not \ell +\not p  +\not k  -m}\frac{1}{\not \ell +\not p  +m}\gamma_{5}%
\]%
\begin{align*}
&  \lim_{n\rightarrow4}\left(  \Gamma_{\ref{bm-o3phi2}\left(  b\right)  }%
^{R5}-\Gamma_{\ref{bm-o3phi2}\left(  b\right)  }^{BM}\right) \\
&  =-if^{3}\int\frac{d^{n}\ell}{\left(  2\pi\right)  ^{n}}\frac{1}{\left(
\ell^{2}-m^{2}\right)  }\left(  \frac{1}{\left(  \ell^{2}-m^{2}\right)
}\gamma_{5}+\frac{1}{\not \ell -m}\gamma_{5}\frac{1}{\not \ell -m}\right) \\
&  =-if^{3}\int\frac{d^{n}\ell}{\left(  2\pi\right)  ^{n}}\frac{1}{\left(
\ell^{2}-m^{2}\right)  ^{2}}\left(  \gamma_{5}+\frac{1}{\left(  \ell^{2}%
-m^{2}\right)  }\not \ell \gamma_{5}\not \ell \right) \\
&  =-2if^{3}\int\frac{d^{n}\ell}{\left(  2\pi\right)  ^{n}}\frac{\ell_{\Delta
}^{2}}{\left(  \ell^{2}-m^{2}\right)  ^{3}}\gamma_{5}=-\frac{1}{\left(
4\pi\right)  ^{2}}f^{3}\gamma_{5}%
\end{align*}

\noindent
Diagram $\left(  c\right)  $%
\[
\Gamma_{\ref{bm-o3phi2}\left(  c\right)  }^{BM}=-f^{3}\int\frac{d^{n}\ell
}{\left(  2\pi\right)  ^{n}}D\left(  \phi_{2},\phi_{2};\ell\right)  \gamma
_{5}\frac{1}{\not \ell +\not p  +\not k  -m}\gamma_{5}\frac{1}{\not \ell
+\not p  -m}\gamma_{5}%
\]%
\[
\Gamma_{\ref{bm-o3phi2}\left(  c\right)  }^{R5}=f^{3}\int\frac{d^{n}\ell
}{\left(  2\pi\right)  ^{n}}D\left(  \phi_{2},\phi_{2};\ell\right)  \frac
{1}{\not \ell +\not p  +\not k  +m}\frac{1}{\not \ell +\not p  -m}\gamma_{5}%
\]%
\begin{align*}
&  \lim_{n\rightarrow4}\left(  \Gamma_{\ref{bm-o3phi2}\left(  c\right)  }%
^{R5}-\Gamma_{\ref{bm-o3phi2}\left(  c\right)  }^{BM}\right) \\
&  =if^{3}\int\frac{d^{n}\ell}{\left(  2\pi\right)  ^{n}}\frac{1}{\left(
\ell^{2}-m^{2}\right)  ^{2}}\left(  \gamma_{5}+\frac{1}{\left(  \ell^{2}%
-m^{2}\right)  }\gamma_{5}\not \ell \gamma_{5}\not \ell \gamma_{5}\right) \\
&  =2if^{3}\int\frac{d^{n}\ell}{\left(  2\pi\right)  ^{n}}\frac{\ell_{\Delta
}^{2}}{\left(  \ell^{2}-m^{2}\right)  ^{3}}\gamma_{5}=\frac{1}{\left(
4\pi\right)  ^{2}}f^{3}\gamma_{5}%
\end{align*}

\noindent
Diagram $\left(  d\right)  $%
\[
\Gamma_{\ref{bm-o3phi2}\left(  d\right)  }^{BM}=-ifg^{2}\int\frac{d^{n}\ell
}{\left(  2\pi\right)  ^{n}}D\left(  H,H;\ell\right)  \left(  k-\ell\right)
^{\mu}D\left(  A_{\mu},A_{\nu};\ell+k\right)  \frac{1}{\not \ell +\not p
+\not k  -m}R\gamma^{\nu}L
\]%
\[
\Gamma_{\ref{bm-o3phi2}\left(  d\right)  }^{R5}=-ifg^{2}\int\frac{d^{n}\ell
}{\left(  2\pi\right)  ^{n}}D\left(  H,H;\ell\right)  \left(  k-\ell\right)
^{\mu}D\left(  A_{\mu},A_{\nu};\ell+k\right)  \frac{1}{\not \ell +\not p
+\not k  -m}\gamma^{\nu}L
\]%
\begin{align*}
&  \lim_{n\rightarrow4}\left(  \Gamma_{\ref{bm-o3phi2}\left(  d\right)  }%
^{R5}-\Gamma_{\ref{bm-o3phi2}\left(  d\right)  }^{BM}\right) \\
&  =-ifg\int\frac{d^{n}\ell}{\left(  2\pi\right)  ^{n}}D\left(  H,H;\ell
\right)  g\left(  k-\ell\right)  ^{\mu}D\left(  A_{\mu},A_{\nu};\ell+k\right)
\frac{1}{\not \ell -m}\left(  \gamma^{\nu}L-R\gamma^{\nu}L\right) \\
&  =ifg^{2}\int\frac{d^{n}\ell}{\left(  2\pi\right)  ^{n}}\frac{\alpha
\ell_{\Delta}^{2}}{\left(  \ell^{2}-m^{2}\right)  ^{3}}L=\frac{1}{\left(
4\pi\right)  ^{2}}\frac{1}{2}\alpha fg^{2}L
\end{align*}

\noindent
Diagram $\left(  e\right)  $%
\[
\Gamma_{\ref{bm-o3phi2}\left(  e\right)  }^{BM}=-igf\int\frac{d^{n}\ell
}{\left(  2\pi\right)  ^{n}}D\left(  A_{\mu},A_{\nu};\ell\right)  g\left(
\ell+2k\right)  ^{\nu}D\left(  H,H;\ell+k\right)  R\gamma^{\mu}L\frac
{1}{\not \ell +\not p  +\not k  -m}%
\]%
\[
\Gamma_{\ref{bm-o3phi2}\left(  e\right)  }^{R5}=-igf\int\frac{d^{n}\ell
}{\left(  2\pi\right)  ^{n}}D\left(  A_{\mu},A_{\nu};\ell\right)  g\left(
\ell+2k\right)  ^{\nu}D\left(  H,H;\ell+k\right)  \gamma^{\mu}\frac{\left(
\not \ell +\not p  +\not k  \right)  R+mL}{\left(  \ell+p+k\right)  ^{2}%
-m^{2}}%
\]%
\begin{align*}
&  \lim_{n\rightarrow4}\left(  \Gamma_{\ref{bm-o3phi2}\left(  e\right)  }%
^{R5}-\Gamma_{\ref{bm-o3phi2}\left(  e\right)  }^{BM}\right) \\
&  =-igf\int\frac{d^{n}\ell}{\left(  2\pi\right)  ^{n}}D\left(  A_{\mu}%
,A_{\nu};\ell\right)  g\ell^{\nu}D\left(  H,H;\ell\right)  \left(  \gamma
^{\mu}\frac{\not \ell R}{\ell^{2}-m^{2}}-R\gamma^{\mu}L\frac{1}{\not \ell
-m}\right) \\
&  =-ig^{2}f\int\frac{d^{n}\ell}{\left(  2\pi\right)  ^{n}}\frac{\alpha
\ell_{\Delta}^{2}}{\left(  \ell^{2}-m^{2}\right)  ^{3}}R=-\frac{1}{\left(
4\pi\right)  ^{2}}\frac{\alpha g^{2}f}{2}R
\end{align*}

\noindent
Summary

The amplitudes and finite counter terms due to diagrams in Figure
$\ref{bm-o3phi2}$ are tabulated in Table \ref{ct-o3phi2}.

$\ $%
\begin{table}[htp] \centering
\begin{tabular}
[c]{|l|l|l|}\hline
Figure & $\left(  4\pi\right)  ^{2}\times\left(  \Gamma^{R5}-\Gamma
^{BM}\right)  $ & $\left(  4\pi\right)  ^{2}\times$ Counter Term\\\hline
$\ref{bm-o3phi2}\left(  a\right)  $ & $0$ & $0$\\
$\ref{bm-o3phi2}\left(  b\right)  $ & $-f^{3}\gamma_{5}$ & $if^{3}\bar{\psi
}\phi_{2}\gamma_{5}\psi$\\
$\ref{bm-o3phi2}\left(  c\right)  $ & $f^{3}\gamma_{5}$ & $-if^{3}\bar{\psi
}\phi_{2}\gamma_{5}\psi$\\
$\ref{bm-o3phi2}\left(  d\right)  $ & $\frac{\alpha g^{2}f}{2}L$ &
$-i\frac{\alpha g^{2}f}{2}\bar{\psi}\phi_{2}L\psi$\\
$\ref{bm-o3phi2}\left(  e\right)  $ & $-\frac{\alpha g^{2}f}{2}R$ &
$i\frac{\alpha g^{2}f}{2}\bar{\psi}\phi_{2}R\psi$\\\hline
\end{tabular}
\caption{Counter terms due to diagrams in Figure \ref{bm-o3phi2}\label{ct-o3phi2}}\label{tb4}%
\end{table}%

\begin{figure}
[tbh]
\begin{center}
\includegraphics[
height=1.0706in,
width=4.8326in
]%
{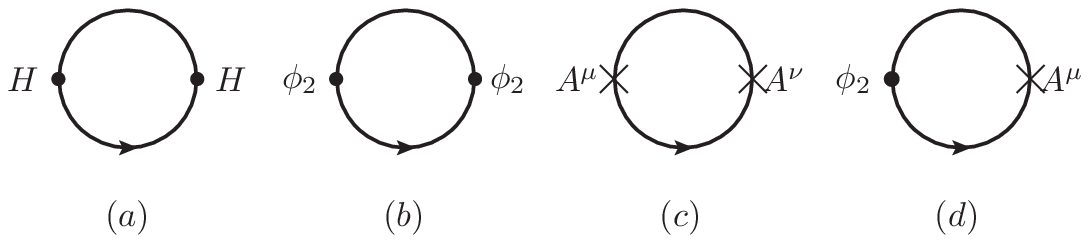}%
\caption{Diagrams for 2-point 1PI}%
\label{bm-c2}%
\end{center}
\end{figure}

\subsubsection{Figure $\ref{bm-c2}$: One-Fermion-Loop $2$-Point 1PI}%

\noindent
Diagram $\left(  a\right)  $

No $\gamma_{5}$ occurs in the amplitude. Thus,%
\[
\Gamma_{\ref{bm-c2}\left(  a\right)  }^{R5}-\Gamma_{\ref{bm-c2}\left(
a\right)  }^{BM}=0
\]

\noindent
Diagram $\left(  b\right)  $%
\[
\Gamma_{\ref{bm-c2}\left(  b\right)  }^{BM}=-f^{2}tr\int\frac{d^{n}\ell
}{\left(  2\pi\right)  ^{n}}\frac{i}{\not \ell -m}\gamma_{5}\frac
{i}{\not \ell -\not p  -m}\gamma_{5}%
\]%
\[
\Gamma_{\ref{bm-c2}\left(  b\right)  }^{R5}=f^{2}tr\int\frac{d^{n}\ell
}{\left(  2\pi\right)  ^{n}}\frac{i}{\not \ell -m}\frac{i}{\not \ell -\not p
+m}%
\]%
\begin{align*}
&  \lim_{n\rightarrow4}\left(  \Gamma_{\ref{bm-c2}\left(  b\right)  }%
^{R5}-\Gamma_{\ref{bm-c2}\left(  b\right)  }^{BM}\right) \\
&  =-f^{2}tr\int\frac{d^{n}\ell}{\left(  2\pi\right)  ^{n}}\frac{1}%
{\not \ell -m}\left(  \frac{1}{\not \ell -\not p  +m}+\gamma_{5}\frac
{1}{\not \ell -\not p  -m}\gamma_{5}\right) \\
&  =-f^{2}\int\frac{d^{n}\ell}{\left(  2\pi\right)  ^{n}}\frac{tr\left[
\left(  \not \ell +m\right)  2\not \ell _{\Delta}\right]  }{\left(  \ell
^{2}-m^{2}\right)  \left(  \left(  \ell-p\right)  ^{2}-m^{2}\right)  }\\
&  =-8f^{2}\int\frac{d^{n}\ell}{\left(  2\pi\right)  ^{n}}\ell_{\Delta}%
^{2}\left(  \frac{1}{\left(  \ell^{2}-m^{2}\right)  ^{2}}+\frac{2\ell\cdot
p-p^{2}}{\left(  \ell^{2}-m^{2}\right)  ^{3}}+\frac{\left(  2\ell\cdot
p-p^{2}\right)  ^{2}}{\left(  \ell^{2}-m^{2}\right)  ^{4}}\right) \\
&  =\frac{1}{\left(  4\pi\right)  ^{2}}if^{2}\left(  8m^{2}-\frac{4}{3}%
p^{2}\right)
\end{align*}

\noindent
Diagram $\left(  c\right)  $%
\[
\Gamma_{\ref{bm-c2}\left(  c\right)  }^{BM}=-g^{2}tr\int\frac{d^{n}\ell
}{\left(  2\pi\right)  ^{n}}\frac{1}{\not \ell -m}R\gamma^{\mu}L\frac
{1}{\not \ell -\not p  -m}R\gamma^{\nu}L
\]%
\[
\Gamma_{\ref{bm-c2}\left(  c\right)  }^{R5}=-g^{2}tr\int\frac{d^{n}\ell
}{\left(  2\pi\right)  ^{n}}\frac{1}{\not \ell -m}\gamma^{\mu}\frac
{\not \ell -\not p  }{\left(  \ell-p\right)  ^{2}-m^{2}}\gamma^{\nu}L
\]%
\begin{align}
&  \lim_{n\rightarrow4}\left(  \Gamma_{\ref{bm-c2}\left(  c\right)  }%
^{R5}-\Gamma_{\ref{bm-c2}\left(  c\right)  }^{BM}\right) \label{fc2c}\\
&  =-g^{2}tr\int\frac{d^{n}\ell}{\left(  2\pi\right)  ^{n}}\frac{1}%
{\not \ell -m}\gamma^{\mu}\left(  \frac{\not \ell -L\not \ell R}{\left(
\ell-p\right)  ^{2}-m^{2}}\right)  \gamma^{\nu}L\nonumber\\
&  =2g^{\mu\nu}g^{2}\int\frac{d^{n}\ell}{\left(  2\pi\right)  ^{n}}%
\ell_{\Delta}^{2}\left(  \frac{1}{\left(  \ell^{2}-m^{2}\right)  ^{2}}%
+\frac{2\ell\cdot p-p^{2}}{\left(  \ell^{2}-m^{2}\right)  ^{3}}+\frac{\left(
2\ell\cdot p-p^{2}\right)  ^{2}}{\left(  \ell^{2}-m^{2}\right)  ^{4}}\right)
\nonumber\\
&  =\frac{1}{\left(  4\pi\right)  ^{2}}ig^{\mu\nu}g^{2}\left(  -2m^{2}%
+\frac{1}{3}p^{2}\right) \nonumber
\end{align}

\noindent
Diagram $\left(  d\right)  $%
\[
\Gamma_{\ref{bm-c2}\left(  d\right)  }^{BM}=-ifg\int\frac{d^{n}\ell}{\left(
2\pi\right)  ^{n}}tr\left(  \frac{1}{\not \ell -\not p  -m}\gamma^{5}\frac
{1}{\not \ell -m}R\gamma^{\mu}L\right)
\]%
\[
\Gamma_{\ref{bm-c2}\left(  d\right)  }^{R5}=-ifg\int\frac{d^{n}\ell}{\left(
2\pi\right)  ^{n}}tr\left(  \frac{1}{\not \ell -\not p  -m}\frac
{1}{-\not \ell -m}\gamma^{\mu}L\right)
\]%
\begin{align*}
&  \lim_{n\rightarrow4}\left(  \Gamma_{\ref{bm-c2}\left(  d\right)  }%
^{R5}-\Gamma_{\ref{bm-c2}\left(  d\right)  }^{BM}\right) \\
&  =ifg\int\frac{d^{n}\ell}{\left(  2\pi\right)  ^{n}}tr\left(  \frac
{1}{\not \ell -\not p  -m}\left(  \frac{1}{\not \ell +m}+\gamma^{5}\frac
{1}{\not \ell -m}R\right)  \gamma^{\mu}L\right) \\
&  =ifg\int\frac{d^{n}\ell}{\left(  2\pi\right)  ^{n}}tr\left(  \frac{2\left(
\not \ell -\not p  +m\right)  \not \ell _{\Delta}\gamma^{\mu}L}{\left(
\left(  \ell-p\right)  ^{2}-m^{2}\right)  \left(  \ell^{2}-m^{2}\right)
}\right)  =0
\end{align*}

\noindent
Summary

The amplitudes and finite counter terms due to diagrams in Figure
$\ref{bm-c2}$ are tabulated in Table \ref{ct-c2}.%

\begin{table}[htp] \centering
\begin{tabular}
[c]{|l|l|l|}\hline
Figure & $\left(  4\pi\right)  ^{2}\times\left(  \Gamma^{R5}-\Gamma
^{BM}\right)  $ & $\left(  4\pi\right)  ^{2}\times$ Counter Term\\\hline
$\ref{bm-c2}\left(  a\right)  $ & $0$ & $0$\\
$\ref{bm-c2}\left(  b\right)  $ & $if^{2}\left(  8m^{2}-\frac{4}{3}%
p^{2}\right)  $ & $4m^{2}f^{2}\left(  \phi_{2}\right)  ^{2}-\frac{2}{3}%
f^{2}\left(  \partial_{\mu}\phi_{2}\right)  \left(  \partial^{\mu}\phi
_{2}\right)  $\\
$\ref{bm-c2}\left(  c\right)  $ & $ig^{\mu\nu}g^{2}\left(  -2m^{2}+\frac{1}%
{3}p^{2}\right)  $ & $-g^{2}m^{2}A^{2}+\frac{1}{6}g^{2}\left(  \partial_{\mu
}A_{\nu}\right)  \left(  \partial^{\mu}A^{\nu}\right)  $\\
$\ref{bm-c2}\left(  d\right)  $ & $0$ & $0$\\\hline
\end{tabular}
\caption{Counter terms due to diagrams in Figure \ref{bm-c2}\label{ct-c2}}\label{tb5}%
\end{table}%

\begin{figure}
[tbh]
\begin{center}
\includegraphics[
height=1.1831in,
width=5.9456in
]%
{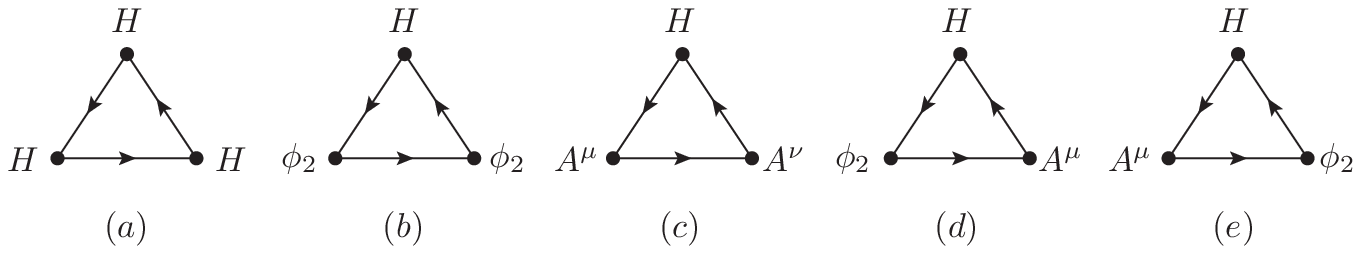}%
\caption{Diagrams for 3-point 1PI}%
\label{bm-c3}%
\end{center}
\end{figure}

\subsubsection{Figure $\ref{bm-c3}$: One-Fermion-Loop $3$-Point 1PI}%

\noindent
Diagram $\left(  a\right)  $

No $\gamma_{5}$ is involved and therefore there is no finite counter term.%

\noindent
Diagram $\left(  b\right)  $%
\[
\Gamma_{\ref{bm-c3}\left(  b\right)  }^{BM}=f^{3}tr\int\frac{d^{n}\ell
}{\left(  2\pi\right)  ^{n}}\frac{1}{\not \ell -m}\frac{1}{\not \ell -\not p
_{1}-m}\gamma_{5}\frac{1}{\not \ell -\not p  _{1}-\not p  _{2}-m}\gamma_{5}%
\]%
\[
\Gamma_{\ref{bm-c3}\left(  b\right)  }^{R5}=-f^{3}tr\int\frac{d^{n}\ell
}{\left(  2\pi\right)  ^{n}}\frac{1}{\not \ell -m}\frac{1}{\not \ell -\not p
_{1}-m}\frac{1}{\not \ell -\not p  _{1}-\not p  _{2}+m}%
\]%
\begin{align*}
&  \lim_{n\rightarrow4}\left(  \Gamma_{\ref{bm-c3}\left(  b\right)  }%
^{R5}-\Gamma_{\ref{bm-c3}\left(  b\right)  }^{BM}\right) \\
&  =f^{3}tr\int\frac{d^{n}\ell}{\left(  2\pi\right)  ^{n}}\frac{1}%
{\not \ell -m}\frac{1}{\not \ell -\not p  _{1}-m}\frac{\left(  -\not \ell
+\not p  _{1}+\not p  _{2}+m\right)  -\gamma_{5}\left(  \not \ell -\not p
_{1}-\not p  _{2}+m\right)  \gamma_{5}}{\left(  \not \ell -\not p
_{1}-\not p  _{2}\right)  ^{2}-m^{2}}\\
&  =-2f^{3}tr\int\frac{d^{n}\ell}{\left(  2\pi\right)  ^{n}}\frac{\left(
\not \ell +m\right)  \left(  \not \ell -\not p  _{1}+m\right)  \not \ell
_{\Delta}}{\left(  \ell^{2}-m^{2}\right)  \left(  \left(  \ell-p_{1}\right)
^{2}-m^{2}\right)  \left(  \left(  \ell-p_{1}-p_{2}\right)  ^{2}-m^{2}\right)
}\\
&  =-2f^{3}tr\int\frac{d^{n}\ell}{\left(  2\pi\right)  ^{n}}\frac{\left(
\not p  _{1}+2m\right)  \ell_{\Delta}^{2}}{\left(  \ell^{2}-m^{2}\right)
^{3}}=\frac{1}{\left(  4\pi\right)  ^{2}}8if^{3}m
\end{align*}
There is another diagram corresponding to the exchange of the two $\phi_{2}$
fields or the reverse of fermion-loop direction which also yields the same
amplitude$\frac{1}{\left(  4\pi\right)  ^{2}}8if^{3}m$.%

\noindent
Diagram $\left(  c\right)  $%
\[
\Gamma_{\ref{bm-c3}\left(  c\right)  }^{BM}=-fg^{2}tr\int\frac{d^{n}\ell
}{\left(  2\pi\right)  ^{n}}\frac{1}{\not \ell -m}\frac{1}{\not \ell -\not p
_{1}-m}R\gamma^{\mu}L\frac{1}{\not \ell -\not p  _{1}-\not p  _{2}-m}%
R\gamma^{\nu}L
\]%
\[
\Gamma_{\ref{bm-c3}\left(  c\right)  }^{R5}=-fg^{2}tr\int\frac{d^{n}\ell
}{\left(  2\pi\right)  ^{n}}\frac{1}{\not \ell -m}\frac{1}{\not \ell -\not p
_{1}-m}\gamma^{\mu}\frac{\not \ell -\not p  _{1}-\not p  _{2}}{\left(
\not \ell -\not p  _{1}-\not p  _{2}\right)  ^{2}-m^{2}}\gamma^{\nu}L
\]%
\begin{align*}
&  \lim_{n\rightarrow4}\left(  \Gamma_{\ref{bm-c3}\left(  c\right)  }%
^{R5}-\Gamma_{\ref{bm-c3}\left(  c\right)  }^{BM}\right) \\
&  =-fg^{2}\int\frac{d^{n}\ell}{\left(  2\pi\right)  ^{n}}\frac{tr\left[
\left(  \not \ell +m\right)  \left(  \not \ell -\not p  _{1}+m\right)
\gamma^{\mu}\not \ell _{\Delta}\gamma^{\nu}L\right]  }{\left(  \ell^{2}%
-m^{2}\right)  \left(  \left(  \ell-p_{1}\right)  ^{2}-m^{2}\right)  \left(
\left(  \ell-p_{1}-p_{2}\right)  ^{2}-m^{2}\right)  }\\
&  =4fg^{2}\int\frac{d^{n}\ell}{\left(  2\pi\right)  ^{n}}\frac{\ell_{\Delta
}^{2}mg^{\mu\nu}}{\left(  \ell^{2}-m^{2}\right)  ^{3}}=-\frac{1}{\left(
4\pi\right)  ^{2}}2ifg^{2}mg^{\mu\nu}%
\end{align*}
Interchanging\ the two external $A$ fields gives us another topologically
different diagram whose amplitude is also equal to $-\frac{1}{\left(
4\pi\right)  ^{2}}2ig^{2}fmg^{\mu\nu}$.%

\noindent
Diagram $\left(  d\right)  $%
\[
\Gamma_{\ref{bm-c3}\left(  d\right)  }^{BM}=-if^{2}gtr\int\frac{d^{n}\ell
}{\left(  2\pi\right)  ^{n}}\frac{1}{\not \ell -m}\frac{1}{\not \ell -\not p
_{1}-m}\gamma_{5}\frac{1}{\not \ell -\not p  _{1}-\not p  _{2}-m}R\gamma^{\nu
}L
\]%
\[
\Gamma_{\ref{bm-c3}\left(  d\right)  }^{R5}=if^{2}gtr\int\frac{d^{n}\ell
}{\left(  2\pi\right)  ^{n}}\frac{1}{\not \ell -m}\frac{1}{\not \ell -\not p
_{1}-m}\frac{1}{\not \ell -\not p  _{1}-\not p  _{2}+m}R\gamma^{\nu}L
\]%
\begin{align*}
&  \lim_{n\rightarrow4}\left(  \Gamma_{\ref{bm-c3}\left(  d\right)  }%
^{R5}-\Gamma_{\ref{bm-c3}\left(  d\right)  }^{BM}\right) \\
&  =if^{2}g\int\frac{d^{n}\ell}{\left(  2\pi\right)  ^{n}}\frac{2tr\left[
\left(  \not \ell +m\right)  \left(  \not \ell -\not p  _{1}+m\right)
\not \ell _{\Delta}\gamma^{\nu}L\right]  }{\left(  \ell^{2}-m^{2}\right)
\left(  \left(  \ell-p_{1}\right)  ^{2}-m^{2}\right)  \left(  \left(
\ell-p_{1}-p_{2}\right)  ^{2}-m^{2}\right)  }\\
&  =if^{2}g\int\frac{d^{n}\ell}{\left(  2\pi\right)  ^{n}}\frac{\ell_{\Delta
}^{2}}{\left(  \ell^{2}-m^{2}\right)  ^{3}}tr\left[  \not p  _{1}\gamma^{\nu
}\right]  =\frac{1}{\left(  4\pi\right)  ^{2}}2f^{2}gp_{1}^{\nu}%
\end{align*}

\noindent
Diagram $\left(  e\right)  $%
\[
\Gamma_{\ref{bm-c3}\left(  e\right)  }^{BM}=-if^{2}gtr\int\frac{d^{n}\ell
}{\left(  2\pi\right)  ^{n}}\frac{1}{\not \ell -m}\frac{1}{\not \ell -\not p
_{1}-m}R\gamma^{\nu}L\frac{1}{\not \ell -\not p  _{1}-\not p  _{2}-m}%
\gamma_{5}%
\]%
\[
\Gamma_{\ref{bm-c3}\left(  e\right)  }^{R5}=-if^{2}gtr\int\frac{d^{n}\ell
}{\left(  2\pi\right)  ^{n}}\frac{1}{\not \ell -m}\frac{1}{\not \ell -\not p
_{1}-m}\gamma^{\nu}\frac{\left(  \not \ell -\not p  _{1}-\not p  _{2}\right)
R-mL}{\left(  \left(  \ell-p_{1}-p_{2}\right)  ^{2}-m^{2}\right)  }%
\]%
\begin{align*}
&  \lim_{n\rightarrow4}\left(  \Gamma_{\ref{bm-c3}\left(  e\right)  }%
^{R5}-\Gamma_{\ref{bm-c3}\left(  e\right)  }^{BM}\right) \\
&  -if^{2}g\int\frac{d^{n}\ell}{\left(  2\pi\right)  ^{n}}\frac{tr\left[
\left(  \not \ell +m\right)  \left(  \not \ell -\not p  _{1}+m\right)
\gamma^{\nu}\not \ell _{\Delta}\right]  }{\left(  \ell^{2}-m^{2}\right)
\left(  \left(  \ell-p_{1}\right)  ^{2}-m^{2}\right)  \left(  \left(
\ell-p_{1}-p_{2}\right)  ^{2}-m^{2}\right)  }\\
&  =if^{2}g\int\frac{d^{n}\ell}{\left(  2\pi\right)  ^{n}}\frac{\ell_{\Delta
}^{2}}{\left(  \ell^{2}-m^{2}\right)  ^{3}}tr\left[  \not p  _{1}\gamma^{\nu
}\right]  =\frac{1}{\left(  4\pi\right)  ^{2}}2f^{2}gp_{1}^{\nu}%
\end{align*}

\noindent
Summary

The amplitudes and finite counter terms due to diagrams in Figure
$\ref{bm-c3}$ are tabulated in Table \ref{ct-c3}. Note the column
"Multiplicity" indicates the combinatorial factor that needs to be multiplied.

$\ $%
\begin{table}[htp] \centering
\begin{tabular}
[c]{|l|l|l|l|}\hline
Figure & $\left(  4\pi\right)  ^{2}\times\left(  \Gamma^{R5}-\Gamma
^{BM}\right)  $ & Multiplicity & $\left(  4\pi\right)  ^{2}\times$ Counter
Term\\\hline
$\ref{bm-c3}\left(  a\right)  $ & $0$ & $2$ & $0$\\
$\ref{bm-c3}\left(  b\right)  $ & $8if^{3}m$ & $2$ & $8f^{3}mH\left(  \phi
_{2}\right)  ^{2}$\\
$\ref{bm-c3}\left(  c\right)  $ & $-2ifg^{2}mg^{\mu\nu}$ & $2$ &
$-2fg^{2}mHA^{2}$\\
$\ref{bm-c3}\left(  d\right)  $ & $2f^{2}gp_{1}^{\nu}$ & $1$ & $2f^{2}%
g\phi_{2}\left(  \partial_{\mu}H\right)  A^{\mu}$\\
$\ref{bm-c3}\left(  e\right)  $ & $2f^{2}gp_{1}^{\nu}$ & $1$ & $2f^{2}%
g\phi_{2}\left(  \partial_{\mu}H\right)  A^{\mu}$\\\hline
\end{tabular}
\caption{Counter terms due to diagrams in Figure \ref{bm-c3}\label{ct-c3}}\label{tb6}%
\end{table}%

\begin{figure}
[tbh]
\begin{center}
\includegraphics[
height=2.0176in,
width=5.2511in
]%
{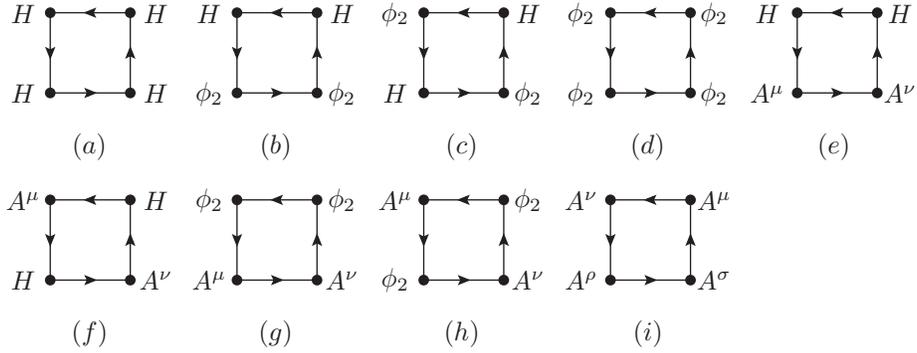}%
\caption{Diagrams for 4-point 1PI}%
\label{bm-c4}%
\end{center}
\end{figure}

\subsubsection{Figure $\ref{bm-c4}$: One-Fermion-Loop $4$-Point 1PI}%

Only $T_{0}$ order terms may be divergent. For the sake of simplicity, we may
assume all the external momenta are zero.%

\noindent
Diagram $\left(  a\right)  $

No $\gamma_{5}$ is involved and therefore diagram $\left(  a\right)  $
generates no finite counter term.%

\noindent
Diagram $\left(  b\right)  $%
\[
\Gamma_{\ref{bm-c4}\left(  b\right)  }^{BM}=f^{4}tr\int\frac{d^{n}\ell
}{\left(  2\pi\right)  ^{n}}\frac{1}{\not \ell -m}\frac{1}{\not \ell -m}%
\frac{1}{\not \ell -m}\gamma_{5}\frac{1}{\not \ell -m}\gamma_{5}%
\]%
\[
\Gamma_{\ref{bm-c4}\left(  b\right)  }^{R5}=f^{4}tr\int\frac{d^{n}\ell
}{\left(  2\pi\right)  ^{n}}\frac{1}{\not \ell -m}\frac{1}{\not \ell -m}%
\frac{1}{\not \ell -m}\frac{-1}{\not \ell +m}%
\]%
\begin{align*}
\lim_{n\rightarrow4}\left(  \Gamma_{\ref{bm-c4}\left(  b\right)  }^{R5}%
-\Gamma_{\ref{bm-c4}\left(  b\right)  }^{BM}\right)   &  =f^{4}\lim
_{n\rightarrow4}tr\int\frac{d^{n}\ell}{\left(  2\pi\right)  ^{n}}%
\frac{\not \ell \not \ell \not \ell }{\left(  \ell^{2}-m^{2}\right)  ^{4}%
}\left(  -\not \ell -\gamma_{5}\not \ell \gamma_{5}\right) \\
&  =-2f^{4}\lim_{n\rightarrow4}\int\frac{d^{n}\ell}{\left(  2\pi\right)  ^{n}%
}\frac{tr\left[  \ell^{2}\ell_{\Delta}^{2}\right]  }{\left(  \ell^{2}%
-m^{2}\right)  ^{4}}=\frac{1}{\left(  4\pi\right)  ^{2}}4if^{4}%
\end{align*}
Exchanging the two $H$ and the two $\phi_{2}$ gives a total counter-term
amplitude of$\frac{1}{\left(  4\pi\right)  ^{2}}$ $4\times4if^{4}=\frac
{1}{\left(  4\pi\right)  ^{2}}16if^{4}$.%

\noindent
Diagram $\left(  c\right)  $%
\[
\Gamma_{\ref{bm-c4}\left(  c\right)  }^{BM}=f^{4}tr\int\frac{d^{n}\ell
}{\left(  2\pi\right)  ^{n}}\frac{1}{\not \ell -m}\frac{1}{\not \ell -m}%
\gamma_{5}\frac{1}{\not \ell -m}\frac{1}{\not \ell -m}\gamma_{5}%
\]%
\[
\Gamma_{\ref{bm-c4}\left(  c\right)  }^{R5}=f^{4}tr\int\frac{d^{n}\ell
}{\left(  2\pi\right)  ^{n}}\frac{1}{\not \ell -m}\frac{1}{\not \ell -m}%
\frac{1}{\not \ell +m}\frac{1}{\not \ell +m}%
\]%
\[
\lim_{n\rightarrow4}\left(  \Gamma_{\ref{bm-c4}\left(  c\right)  }^{R5}%
-\Gamma_{\ref{bm-c4}\left(  c\right)  }^{BM}\right)  =f^{4}tr\int\frac
{d^{n}\ell}{\left(  2\pi\right)  ^{n}}\frac{\not \ell \not \ell
\not \ell \not \ell -\not \ell \not \ell \gamma_{5}\not \ell \not \ell
\gamma_{5}}{\left(  \ell^{2}-m^{2}\right)  ^{4}}=0
\]

\noindent
Diagram $\left(  d\right)  $%
\[
\Gamma_{\ref{bm-c4}\left(  d\right)  }^{BM}=-f^{4}tr\int\frac{d^{n}\ell
}{\left(  2\pi\right)  ^{n}}\frac{1}{\not \ell -m}\gamma_{5}\frac
{1}{\not \ell -m}\gamma_{5}\frac{1}{\not \ell -m}\gamma_{5}\frac{1}%
{\not \ell -m}\gamma_{5}%
\]%
\[
\Gamma_{\ref{bm-c4}\left(  d\right)  }^{R5}=-f^{4}tr\int\frac{d^{n}\ell
}{\left(  2\pi\right)  ^{n}}\frac{1}{\not \ell -m}\frac{1}{\not \ell +m}%
\frac{1}{\not \ell -m}\frac{1}{\not \ell +m}%
\]%
\begin{align*}
\lim_{n\rightarrow4}\left(  \Gamma_{\ref{bm-c4}\left(  d\right)  }^{R5}%
-\Gamma_{\ref{bm-c4}\left(  d\right)  }^{BM}\right)   &  =-f^{4}\int
\frac{d^{n}\ell}{\left(  2\pi\right)  ^{n}}\frac{tr\left[  \ell^{4}%
-\not \ell \gamma_{5}\not \ell \gamma_{5}\not \ell \gamma_{5}\not \ell
\gamma_{5}\right]  }{\left(  \ell^{2}-m^{2}\right)  ^{4}}\\
&  =-32f^{4}\int\frac{d^{n}\ell}{\left(  2\pi\right)  ^{n}}\frac{\ell_{\Delta
}^{2}\underline{\ell}^{2}}{\left(  \ell^{2}-m^{2}\right)  ^{4}}=\frac
{1}{\left(  4\pi\right)  ^{2}}\frac{32}{3}if^{4}%
\end{align*}
Permutation of the four $\phi_{2}$ on a loop gives a combinatorial factor of
$3!$. The total counter-term amplitude due to this type of diagram is
$3!\times\frac{1}{\left(  4\pi\right)  ^{2}}\frac{32}{3}if^{4}=\frac
{1}{\left(  4\pi\right)  ^{2}}64if^{4}$%

\noindent
Diagram $\left(  e\right)  $%
\[
\Gamma_{\ref{bm-c4}\left(  e\right)  }^{BM}=-f^{2}g^{2}tr\int\frac{d^{n}\ell
}{\left(  2\pi\right)  ^{n}}\frac{1}{\not \ell -m}\frac{1}{\not \ell -m}%
\frac{1}{\not \ell -m}R\gamma^{\mu}L\frac{1}{\not \ell -m}R\gamma^{\nu}L
\]%
\[
\Gamma_{\ref{bm-c4}\left(  e\right)  }^{R5}=-f^{2}g^{2}tr\int\frac{d^{n}\ell
}{\left(  2\pi\right)  ^{n}}\frac{1}{\not \ell -m}\frac{1}{\not \ell -m}%
\frac{1}{\not \ell -m}\gamma^{\mu}\frac{\not \ell }{\left(  \ell^{2}%
-m^{2}\right)  }\gamma^{\nu}L
\]%
\begin{align*}
\lim_{n\rightarrow4}\left(  \Gamma_{\ref{bm-c4}\left(  e\right)  }^{R5}%
-\Gamma_{\ref{bm-c4}\left(  e\right)  }^{BM}\right)   &  =-f^{2}g^{2}\int
\frac{d^{n}\ell}{\left(  2\pi\right)  ^{n}}\frac{tr\left[  \ell^{2}%
\not \ell \gamma^{\mu}\not \ell \gamma^{\nu}L-\ell^{2}\not \ell \gamma^{\mu
}\underline{\not \ell }\gamma^{\nu}L\right]  }{\left(  \ell^{2}-m^{2}\right)
^{4}}\\
&  =-f^{2}g^{2}\int\frac{d^{n}\ell}{\left(  2\pi\right)  ^{n}}\frac{tr\left[
\not \ell \gamma^{\mu}\not \ell _{\Delta}\gamma^{\nu}L\right]  }{\left(
\ell^{2}-m^{2}\right)  ^{3}}=-\frac{1}{\left(  4\pi\right)  ^{2}}if^{2}%
g^{2}g^{\mu\nu}%
\end{align*}
Exchanges of the two $H$ and of the two $A$ multiply the above amplitude by a
factor of $4$. i.e., the total amplitude is $-\frac{1}{\left(  4\pi\right)
^{2}}i4f^{2}g^{2}g^{\mu\nu}$.%

\noindent
Diagram $\left(  f\right)  $%
\[
\Gamma_{\ref{bm-c4}\left(  f\right)  }^{BM}=-f^{2}g^{2}tr\int\frac{d^{n}\ell
}{\left(  2\pi\right)  ^{n}}\frac{1}{\not \ell -m}\frac{1}{\not \ell
-m}R\gamma^{\mu}L\frac{1}{\not \ell -m}\frac{1}{\not \ell -m}R\gamma^{\nu}L
\]%
\[
\Gamma_{\ref{bm-c4}\left(  f\right)  }^{R5}=-f^{2}g^{2}tr\int\frac{d^{n}\ell
}{\left(  2\pi\right)  ^{n}}\frac{1}{\not \ell -m}\frac{1}{\not \ell -m}%
\gamma^{\mu}\frac{2m\not \ell }{\left(  \ell^{2}-m^{2}\right)  ^{2}}%
\gamma^{\nu}L
\]%
\[
\lim_{n\rightarrow4}\left(  \Gamma_{\ref{bm-c4}\left(  f\right)  }^{R5}%
-\Gamma_{\ref{bm-c4}\left(  f\right)  }^{BM}\right)  =0
\]

\noindent
Diagram $\left(  g\right)  $%
\[
\Gamma_{\ref{bm-c4}\left(  g\right)  }^{BM}=f^{2}g^{2}tr\int\frac{d^{n}\ell
}{\left(  2\pi\right)  ^{n}}\frac{1}{\not \ell -m}\gamma_{5}\frac
{1}{\not \ell -m}\gamma_{5}\frac{1}{\not \ell -m}R\gamma^{\mu}L\frac
{1}{\not \ell -m}R\gamma^{\nu}L
\]%
\[
\Gamma_{\ref{bm-c4}\left(  g\right)  }^{R5}=-f^{2}g^{2}tr\int\frac{d^{n}\ell
}{\left(  2\pi\right)  ^{n}}\frac{1}{\not \ell -m}\frac{1}{\not \ell +m}%
\frac{1}{\not \ell -m}\gamma^{\mu}\frac{\not \ell }{\ell^{2}-m^{2}}\gamma
^{\nu}L
\]%
\begin{align*}
&  \lim_{n\rightarrow4}\left(  \Gamma_{\ref{bm-c4}\left(  g\right)  }%
^{R5}-\Gamma_{\ref{bm-c4}\left(  g\right)  }^{BM}\right) \\
&  =-f^{2}g^{2}\int\frac{d^{n}\ell}{\left(  2\pi\right)  ^{n}}\frac{tr\left[
\ell^{2}\not \ell \gamma^{\mu}\not \ell \gamma^{\nu}L+\not \ell \gamma
_{5}\not \ell \gamma_{5}\not \ell R\gamma^{\mu}L\not \ell R\gamma^{\nu
}L\right]  }{\left(  \ell^{2}-m^{2}\right)  ^{4}}\\
&  =\frac{1}{2}f^{2}g^{2}\int\frac{d^{n}\ell}{\left(  2\pi\right)  ^{n}}%
\frac{\ell_{\Delta}^{2}\left(  \ell^{2}+2\underline{\ell}^{2}\right)
}{\left(  \ell^{2}-m^{2}\right)  ^{4}}tr\left[  \gamma^{\mu}\gamma^{\nu
}\right]  =-\frac{1}{\left(  4\pi\right)  ^{2}}i\frac{7}{3}f^{2}g^{2}g^{\mu
\nu}%
\end{align*}
Permuting the two external $\phi_{2}$ and the two external $A$ yields three
additional diagrams each of them contributes the same counter-term amplitude
as the above. The total counter-term amplitude is therefore equal to
$2\times2\times\left(  -\frac{1}{\left(  4\pi\right)  ^{2}}i\frac{7}{3}%
f^{2}g^{2}g^{\mu\nu}\right)  =-\frac{1}{\left(  4\pi\right)  ^{2}}i\frac
{28}{3}f^{2}g^{2}g^{\mu\nu}$%

\noindent
Diagram $\left(  h\right)  $%
\[
\Gamma_{\ref{bm-c4}\left(  h\right)  }^{BM}=f^{2}g^{2}tr\int\frac{d^{n}\ell
}{\left(  2\pi\right)  ^{n}}\frac{1}{\not \ell -m}\gamma_{5}\frac
{1}{\not \ell -m}R\gamma^{\mu}L\frac{1}{\not \ell -m}\gamma_{5}\frac
{1}{\not \ell -m}R\gamma^{\nu}L
\]
Rightmost-$\gamma_{5}$ amplitude vanishes in the $T_{0}$ order.%
\[
\Gamma_{\ref{bm-c4}\left(  h\right)  }^{R5}=0
\]%
\begin{align*}
&  \lim_{n\rightarrow4}\left(  \Gamma_{\ref{bm-c4}\left(  h\right)  }%
^{R5}-\Gamma_{\ref{bm-c4}\left(  h\right)  }^{BM}\right) \\
&  =4f^{2}g^{2}tr\int\frac{d^{n}\ell}{\left(  2\pi\right)  ^{n}}\frac
{-\ell_{\Delta}^{2}\underline{\not \ell }\gamma^{\mu}\underline{\not \ell
}\gamma^{\nu}L}{\left(  \ell^{2}-m^{2}\right)  ^{4}}=4f^{2}g^{2}\frac
{4}{n\left(  n+2\right)  }\int\frac{d^{n}\ell}{\left(  2\pi\right)  ^{n}}%
\frac{\left(  n-4\right)  }{\left(  \ell^{2}-m^{2}\right)  ^{2}}g^{\mu\nu}\\
&  =-\frac{1}{\left(  4\pi\right)  ^{2}}\frac{4}{3}if^{2}g^{2}g^{\mu\nu}%
\end{align*}
By reversing the loop direction, or by exchanging the two external $\phi_{2}$
fields or the two external $A$ fields, we obtain another diagram that also
contributes the same counter-term amplitude as the above. The total
counter-term amplitude is $2\times\left(  -\frac{1}{\left(  4\pi\right)  ^{2}%
}\frac{4}{3}if^{2}g^{2}g^{\mu\nu}\right)  =-\frac{1}{\left(  4\pi\right)
^{2}}\frac{8}{3}if^{2}g^{2}g^{\mu\nu}$.%

\noindent
Diagram $\left(  i\right)  $%
\[
\Gamma_{\ref{bm-c4}\left(  i\right)  }^{BM}=-g^{4}\int\frac{d^{n}\ell}{\left(
2\pi\right)  ^{n}}\frac{tr\left[  R\gamma^{\mu}L\not \ell R\gamma^{\nu
}L\not \ell R\gamma^{\rho}L\not \ell R\gamma^{\sigma}L\right]  }{\left(
\ell^{2}-m^{2}\right)  ^{4}}%
\]%
\[
\Gamma_{\ref{bm-c4}\left(  i\right)  }^{R5}=-g^{4}\int\frac{d^{n}\ell}{\left(
2\pi\right)  ^{n}}\frac{tr\left[  \not \ell \gamma^{\mu}\not \ell \gamma^{\nu
}\not \ell \gamma^{\rho}\not \ell \gamma^{\sigma}L\right]  }{\left(  \ell
^{2}-m^{2}\right)  ^{4}}%
\]%
\begin{align}
&  \lim_{n\rightarrow4}\left(  \Gamma_{\ref{bm-c4}\left(  i\right)  }%
^{R5}-\Gamma_{\ref{bm-c4}\left(  i\right)  }^{BM}\right) \label{fc4i}\\
&  =-g^{4}\int\frac{d^{n}\ell}{\left(  2\pi\right)  ^{n}}\frac{tr\left[
\not \ell \gamma^{\mu}\not \ell \gamma^{\nu}\not \ell \gamma^{\rho}%
\not \ell \gamma^{\sigma}L-\underline{\not \ell }\gamma^{\mu}\underline
{\not \ell }\gamma^{\nu}\underline{\not \ell }\gamma^{\rho}\underline
{\not \ell }\gamma^{\sigma}L\right]  }{\left(  \ell^{2}-m^{2}\right)  ^{4}%
}\nonumber\\
&  =\frac{1}{2}g^{4}\int\frac{d^{n}\ell}{\left(  2\pi\right)  ^{n}}\frac
{\ell_{\Delta}^{2}tr\left[  \underline{\ell}^{2}\left(  -2\gamma^{\mu}%
\gamma^{\nu}\gamma^{\rho}\gamma^{\sigma}+\gamma^{\mu}\gamma^{\sigma}g^{\nu
\rho}+g^{\mu\nu}\gamma^{\rho}\gamma^{\sigma}\right)  -\ell_{\Delta}^{2}%
\gamma^{\mu}\gamma^{\nu}\gamma^{\rho}\gamma^{\sigma}\right]  }{\left(
\ell^{2}-m^{2}\right)  ^{4}}\nonumber\\
&  =\frac{1}{\left(  4\pi\right)  ^{2}}ig^{4}\left(  g^{\mu\nu}g^{\rho\sigma
}-\frac{5}{3}g^{\mu\rho}g^{\nu\sigma}+g^{\mu\sigma}g^{\rho\nu}\right)
\nonumber
\end{align}
The above amplitude is invariant if we reverse the loop direction or make the
interchange $\left(  \mu\leftrightarrow\sigma\right)  $. For the 4-point
$AAAA$ 1PI function, there are in total 6 topologically different diagrams
that may be obtained from Figure $\ref{bm-c4}\left(  i\right)  $ by permuting
the indices $\nu$,$\rho$, and $\sigma$. The total amplitude for $AAAA$ is
equal to%
\begin{align*}
&  \frac{1}{\left(  4\pi\right)  ^{2}}i2g^{4}\left(  g^{\mu\nu}g^{\rho\sigma
}-\frac{5}{3}g^{\mu\rho}g^{\nu\sigma}+g^{\mu\sigma}g^{\rho\nu}+\left(
\rho\leftrightarrow\nu\right)  +\left(  \rho\leftrightarrow\sigma\right)
\right) \\
&  =\frac{1}{\left(  4\pi\right)  ^{2}}i\frac{2}{3}g^{4}\left(  g^{\mu\nu
}g^{\rho\sigma}+g^{\mu\rho}g^{\nu\sigma}+g^{\mu\sigma}g^{\rho\nu}\right)
\end{align*}

\noindent
Summary

The amplitudes and finite counter terms due to diagrams in Figure
$\ref{bm-c4}$ are tabulated in Table \ref{ct-c4}.%

\begin{table}[htbp] \centering
\begin{tabular}
[c]{|l|l|l|l|}\hline
Figure & $\left(  4\pi\right)  ^{2}\times\left(  \Gamma^{R5}-\Gamma
^{BM}\right)  $ & Multiplicity & $\left(  4\pi\right)  ^{2}\times$ Counter
Term\\\hline
$\ref{bm-c4}\left(  a\right)  $ & $0$ & $3!$ & $0$\\
$\ref{bm-c4}\left(  b\right)  $ & $4if^{4}$ & $4$ & $4f^{4}H^{2}\left(
\phi_{2}\right)  ^{2}$\\
$\ref{bm-c4}\left(  c\right)  $ & $0$ & $2$ & $0$\\
$\ref{bm-c4}\left(  d\right)  $ & $\frac{32}{3}if^{4}$ & $3!$ & $\frac{8}%
{3}f^{4}\left(  \phi_{2}\right)  ^{4}$\\
$\ref{bm-c4}\left(  e\right)  $ & $-if^{2}g^{2}g^{\mu\nu}$ & $4$ &
$-f^{2}g^{2}H^{2}A^{2}$\\
$\ref{bm-c4}\left(  f\right)  $ & $0$ & $2$ & $0$\\
$\ref{bm-c4}\left(  g\right)  $ & $-i\frac{7}{3}f^{2}g^{2}g^{\mu\nu}$ & $4$ &
$-\frac{7}{3}f^{2}g^{2}\left(  \phi_{2}\right)  ^{2}A^{2}$\\
$\ref{bm-c4}\left(  h\right)  $ & $-i\frac{4}{3}f^{2}g^{2}g^{\mu\nu}$ & $2$ &
$-\frac{2}{3}f^{2}g^{2}\left(  \phi_{2}\right)  ^{2}A^{2}$\\
$%
\begin{array}
[c]{c}%
\ref{bm-c4}\left(  i\right)  +\left(  \rho\leftrightarrow\nu\right) \\
+\left(  \rho\leftrightarrow\sigma\right)
\end{array}
$ & $i\frac{g^{4}}{3}\left(
\begin{array}
[c]{c}%
g^{\mu\nu}g^{\rho\sigma}+g^{\mu\rho}g^{\nu\sigma}\\
+g^{\mu\sigma}g^{\rho\nu}%
\end{array}
\right)  $ & $2$ & $\frac{1}{12}g^{4}\left(  A^{2}\right)  ^{2}$\\\hline
\end{tabular}
\caption{Counter terms due to diagrams in Figure \ref{bm-c4}\label{ct-c4}}\label{tb7}%
\end{table}%

\section{The Chiral Non-Abelian Gauge Theory\label{apxb}}

\subsection{Feynman Rules\label{nfrs}}

The propagators and vertices used in the 1-loop counter-term calculation for
the chiral Abelian-Higgs theory defined by $\left(  \ref{nbm}\right)  $ are
listed below.

\subsubsection{Propagators:}

\begin{center}%
\[
S\left(  \psi,\bar{\psi};p\right)  :%
\begin{array}
[b]{c}%
\begin{picture}(92,23) (15,-10)
\SetWidth{0.5}
\SetColor{Black}
\Text(60,3)[]{\normalsize{\Black{$p$}}}
\SetWidth{1.0}
\Line
[arrow,arrowpos=0.5,arrowlength=3,arrowwidth=1.2,arrowinset=0.2](90,-7)(30,-7)
\Text(30,3)[]{\normalsize{\Black{$\psi$}}}
\Text(90,3)[]{\normalsize{\Black{$\bar{\psi}$}}}
\end{picture}%
\end{array}
=\frac{i}{\not p  }%
\]%
\[
S\left(  \psi^{\prime},\bar{\psi}^{\prime};p\right)  :%
\begin{array}
[b]{c}%
\begin{picture}(92,23) (15,-10)
\SetWidth{0.5}
\SetColor{Black}
\Text(60,3)[]{\normalsize{\Black{$p$}}}
\SetWidth{1.0}
\Line
[arrow,arrowpos=0.5,arrowlength=3,arrowwidth=1.2,arrowinset=0.2](90,-7)(30,-7)
\Text(30,3)[]{\normalsize{\Black{$\psi^{\prime}$}}}
\Text(90,3)[]{\normalsize{\Black{$\bar{\psi}^{\prime}$}}}
\end{picture}%
\end{array}
=\frac{i}{\not p  }%
\]%
\[
D\left(  A^{a,\mu},A^{b,\nu};k\right)  :%
\begin{array}
[c]{c}%
\begin{picture}(92,36) (15,-10)
\SetWidth{0.5}
\SetColor{Black}
\Photon(30,6)(90,6){2.5}{6}
\Text(60,16)[]{\normalsize{\Black{$k$}}}
\Text(30,-4)[]{\normalsize{\Black{$a,\mu$}}}
\Text(90,-4)[]{\normalsize{\Black{$b,\nu$}}}
\Line
[arrow,arrowpos=1,arrowlength=2.5,arrowwidth=1,arrowinset=0.2](50,16)(40,16)
\end{picture}%
\end{array}
=\frac{-i}{k^{2}}\left(  g^{\mu\nu}+\left(  \alpha-1\right)  \frac{k^{\mu
}k^{\nu}}{k^{2}}\right)  \delta^{ab}%
\]

\end{center}

\subsubsection{Vertex Factors:}%

\[
\bar{\psi}A^{a,\mu}\psi:%
\begin{array}
[c]{c}%
\begin{picture}(62,37) (60,-19)
\SetWidth{0.5}
\SetColor{Black}
\Photon(90,17)(90,-3){2.5}{4}
\Text(105,7)[]{\normalsize{\Black{$A^{a,\mu}$}}}
\Line
[dash,dashsize=0.3,arrow,arrowpos=0.5,arrowlength=2.5,arrowwidth=1,arrowinset=0.2](90,-3)(60,-3)
\Line
[dash,dashsize=0.3,arrow,arrowpos=0.5,arrowlength=2.5,arrowwidth=1,arrowinset=0.2](120,-3)(90,-3)
\Text(80,-13)[]{\normalsize{\Black{$\bar{\psi}$}}}
\Text(100,-13)[]{\normalsize{\Black{$\psi$}}}
\end{picture}%
\end{array}
=-igR\gamma^{\mu}LT_{L}^{a}%
\]%
\[
\bar{\psi}^{\prime}A^{a,\mu}\psi^{\prime}:%
\begin{array}
[c]{c}%
\begin{picture}(62,37) (60,-19)
\SetWidth{0.5}
\SetColor{Black}
\Photon(90,17)(90,-3){2.5}{4}
\Text(105,7)[]{\normalsize{\Black{$A^{a,\mu}$}}}
\Line
[dash,dashsize=0.3,arrow,arrowpos=0.5,arrowlength=2.5,arrowwidth=1,arrowinset=0.2](90,-3)(60,-3)
\Line
[dash,dashsize=0.3,arrow,arrowpos=0.5,arrowlength=2.5,arrowwidth=1,arrowinset=0.2](120,-3)(90,-3)
\Text(80,-13)[]{\normalsize{\Black{$\bar{\psi}^{\prime}$}}}
\Text(100,-13)[]{\normalsize{\Black{$\psi^{\prime}$}}}
\end{picture}%
\end{array}
=-igL\gamma^{\mu}RT_{R}^{a}%
\]%
\begin{align*}
A_{\mu}^{a}A_{\nu}^{b}A_{\rho}^{c}  &  :%
\begin{array}
[c]{c}%
\begin{picture}(96,86) (51,-20)
\SetWidth{0.5}
\SetColor{Black}
\Photon(100,46)(100,26){2.5}{4}
\Text(105,56)[]{\normalsize{\Black{$A^{a,\mu}$}}}
\Photon(100,26)(80,6){2.5}{6}
\Photon(100,26)(120,6){2.5}{6}
\Vertex(100,26){2}
\Text(130,6)[]{\normalsize{\Black{$A^{c,\rho}$}}}
\Text(66,6)[]{\normalsize{\Black{$A^{b,\nu}$}}}
\Text(120,36)[]{\normalsize{\Black{$k_1$}}}
\Text(80,-14)[]{\normalsize{\Black{$k_2$}}}
\Text(120,-14)[]{\normalsize{\Black{$k_3$}}}
\Line
[arrow,arrowpos=1,arrowlength=2.5,arrowwidth=1,arrowinset=0.2](110,46)(110,36)
\Line
[arrow,arrowpos=1,arrowlength=2.5,arrowwidth=1,arrowinset=0.2](80,-4)(90,6)
\Line
[arrow,arrowpos=1,arrowlength=2.5,arrowwidth=1,arrowinset=0.2](120,-4)(110,6)
\end{picture}%
\end{array}
\\
&  =-gC^{abc}\left(  g^{\mu\nu}\left(  k_{1}-k_{2}\right)  ^{\rho}+g^{\nu\rho
}\left(  k_{2}-k_{3}\right)  ^{\mu}+g^{\mu\rho}\left(  k_{3}-k_{1}\right)
^{\nu}\right)
\end{align*}

\subsection{1-Loop Counter Terms}

\begin{figure}
[tbh]
\begin{center}
\includegraphics[
height=0.4964in,
width=0.8475in
]%
{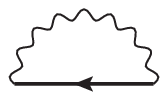}%
\caption{Fermion Self-Energy Diagram}%
\label{bm-no2}%
\end{center}
\end{figure}

\subsubsection{Figure \ref{bm-no2}: Fermion Self-Energy Diagram}%

The fermion field in Figure $\ref{bm-no2}$ can be either $\psi$ or
$\psi^{\prime}$. If the fermion field is $\psi$, the Feynman integral for
Figure $\ref{bm-no2}$ without including the non-Abelian group factor $\sum
_{e}T_{L}^{e}T_{L}^{e}=C_{L}$ is the same as the one for the diagram of
$\left(  \ref{bm-fct1}\right)  $ of which the counter term has been
demonstrated to be $\left(  \ref{fc5}\right)  $ in Sec. \ref{exsc}. If the
fermion field is $\psi^{\prime}$, the counter term can be obtained from that
for the $\psi$ field by replacing $\gamma_{5}$ with \ $-\gamma_{5}$ and the
group factor $C_{L}$ with $\sum_{e}T_{R}^{e}T_{R}^{e}=C_{R}$. The counter term
therefore is equal to
\begin{equation}
-\frac{1}{\left(  4\pi\right)  ^{2}}\frac{1}{3}g^{2}\left(  1+2\alpha\right)
\left(  \bar{\psi}_{L}i\not \partial \psi_{L}C_{L}+\bar{\psi}_{R}^{\prime
}i\not \partial \psi_{R}^{\prime}C_{R}\right)  \label{ct-no2}%
\end{equation}

\begin{figure}
[tbh]
\begin{center}
\includegraphics[
height=1.1346in,
width=1.9418in
]%
{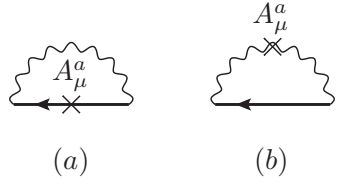}%
\caption{$\bar{\psi}A\psi$ and $\bar{\psi}^{\prime}A\psi^{\prime}$ Vertex
Diagrams }%
\label{bm-no3A}%
\end{center}
\end{figure}

\subsubsection{Figure $\ref{bm-no3A}$: $\bar{\psi}A\psi$ and $\bar{\psi
}^{\prime}A\psi^{\prime}$ Vertex Diagrams}%

\noindent
Diagram $\left(  a\right)  $

The amplitude for this diagram excluding the non-Abelian group factor is the
same as that for Figure $\ref{bm-o3A}\left(  a\right)  $. If $\psi$ is the
fermion field, the non-Abelian group factor is $\sum_{e}T_{L}^{e}T_{L}%
^{a}T_{L}^{e}=C_{L}T_{L}^{a}+iC^{abc}T_{L}^{b}T_{L}^{c}$ and
\begin{equation}
\lim_{n\rightarrow4}\left(  \Gamma_{\psi,\ref{bm-no3A}\left(  a\right)  }%
^{R5}-\Gamma_{\psi,\ref{bm-no3A}\left(  a\right)  }^{BM}\right)  =\frac
{1}{\left(  4\pi\right)  ^{2}}\frac{ig^{3}}{6}\left(  7+5\alpha\right)
\gamma^{\mu}L\left(  C_{L}T_{L}^{a}+iC^{abc}T_{L}^{b}T_{L}^{c}\right)
\label{n3f1}%
\end{equation}
If $\psi^{\prime}$ is the fermion field, the non-Abelian group factor is
$\sum_{e}T_{R}^{e}T_{R}^{a}T_{R}^{e}=C_{R}T_{R}^{a}+iC^{abc}T_{R}^{b}T_{R}%
^{c}$ and
\begin{equation}
\lim_{n\rightarrow4}\left(  \Gamma_{\psi^{\prime},\ref{bm-no3A}\left(
a\right)  }^{R5}-\Gamma_{\psi^{\prime},\ref{bm-no3A}\left(  a\right)  }%
^{BM}\right)  =\frac{1}{\left(  4\pi\right)  ^{2}}\frac{ig^{3}}{6}\left(
7+5\alpha\right)  \gamma^{\mu}R\left(  C_{R}T_{R}^{a}+iC^{abc}T_{R}^{b}%
T_{R}^{c}\right)  \label{n3f2}%
\end{equation}
The counter term that is responsible for the amplitudes of $\left(
\ref{n3f1}\right)  $ and $\left(  \ref{n3f2}\right)  $ is%
\begin{align}
&  \frac{1}{6}g^{3}\left(  7+5\alpha\right)  \bar{\psi}_{L}\not A  \psi
_{L}\left(  C_{L}T_{L}^{a}+iC^{abc}T_{L}^{b}T_{L}^{c}\right)  \label{ct-no3Aa}%
\\
&  +\frac{1}{6}g^{3}\left(  7+5\alpha\right)  \bar{\psi}_{R}\not A  \psi
_{R}\left(  C_{R}T_{R}^{a}+iC^{abc}T_{R}^{b}T_{R}^{c}\right) \nonumber
\end{align}%
\noindent
Diagram $\left(  b\right)  $

If $\psi$ is the fermion field, we have%
\begin{align*}
\Gamma_{\psi,\ref{bm-no3A}\left(  b\right)  }^{BM} &  =-ig^{3}\lim
_{m\rightarrow0}\int\frac{d^{n}\ell}{\left(  2\pi\right)  ^{n}}R\gamma^{\tau
}L\frac{1}{\not \ell-m}R\gamma^{\sigma}LC^{bac}T_{L}^{b}T_{L}^{c}\\
&  \times D_{\tau\nu}\left(  \ell-k_{1}\right)  \left(  2g^{\nu\rho}2\ell
^{\mu}-g^{\mu\nu}\ell^{\rho}-g^{\mu\rho}\ell^{\nu}\right)  D_{\rho\sigma
}\left(  \ell\right)
\end{align*}
and%
\begin{align}
&  \lim_{n\rightarrow4}\left(  \Gamma_{\psi,\ref{bm-no3A}\left(  b\right)
}^{R5}-\Gamma_{\psi,\ref{bm-no3A}\left(  b\right)  }^{BM}\right)
\label{n3f3}\\
&  =-ig^{3}\lim_{m\rightarrow0}\int D_{\tau\nu}\left(  \ell\right)  \left(
2g^{\nu\rho}2\ell^{\mu}-g^{\mu\nu}\ell^{\rho}-g^{\mu\rho}\ell^{\nu}\right)
D_{\rho\sigma}\left(  \ell\right)  \nonumber\\
&  \times\frac{1}{\ell^{2}-m^{2}}\left(  \gamma^{\tau}\not \ell\gamma^{\sigma
}-\underline{\gamma}^{\tau}\underline{\not \ell}\underline{\gamma}^{\sigma
}\right)  LC^{bac}T_{L}^{b}T_{L}^{c}\nonumber\\
&  =\frac{1}{\left(  4\pi\right)  ^{2}}\frac{1}{6}\left(  7+5\alpha\right)
g^{3}\gamma^{\mu}LC^{abc}T_{L}^{b}T_{L}^{c}\nonumber
\end{align}
If $\psi^{\prime}$ is the fermion field, we have%
\begin{align*}
\Gamma_{\psi^{\prime},\ref{bm-no3A}\left(  b\right)  }^{BM} &  =-ig^{3}%
\lim_{m\rightarrow0}\int\frac{d^{n}\ell}{\left(  2\pi\right)  ^{n}}%
L\gamma^{\tau}R\frac{1}{\not \ell-m}L\gamma^{\sigma}RC^{bac}T_{R}^{b}T_{R}%
^{c}\\
&  \times D_{\tau\nu}\left(  \ell-k_{1}\right)  \left(  2g^{\nu\rho}2\ell
^{\mu}-g^{\mu\nu}\ell^{\rho}-g^{\mu\rho}\ell^{\nu}\right)  D_{\rho\sigma
}\left(  \ell\right)
\end{align*}
and%
\begin{align}
&  \lim_{n\rightarrow4}\left(  \Gamma_{\psi^{\prime},\ref{bm-no3A}\left(
b\right)  }^{R5}-\Gamma_{\psi^{\prime},\ref{bm-no3A}\left(  b\right)  }%
^{BM}\right)  \label{n3f4}\\
&  =-ig^{3}\lim_{m\rightarrow0}\int D_{\tau\nu}\left(  \ell\right)  \left(
2g^{\nu\rho}2\ell^{\mu}-g^{\mu\nu}\ell^{\rho}-g^{\mu\rho}\ell^{\nu}\right)
D_{\rho\sigma}\left(  \ell\right)  \nonumber\\
&  \times\frac{1}{\ell^{2}-m^{2}}\left(  \gamma^{\tau}\not \ell\gamma^{\sigma
}-\underline{\gamma}^{\tau}\underline{\not \ell}\underline{\gamma}^{\sigma
}\right)  RC^{bac}T_{R}^{b}T_{R}^{c}\nonumber\\
&  =\frac{1}{\left(  4\pi\right)  ^{2}}\frac{1}{6}\left(  7+5\alpha\right)
g^{3}\gamma^{\mu}RC^{abc}T_{R}^{b}T_{R}^{c}\nonumber
\end{align}
The counter term to generate the amplitudes of $\left(  \ref{n3f3}\right)  $
and $\left(  \ref{n3f4}\right)  $ is%
\begin{equation}
-i\frac{1}{\left(  4\pi\right)  ^{2}}\frac{1}{6}g^{3}\left(  7+5\alpha\right)
\left(  \bar{\psi}_{L}\not A\psi_{L}C^{abc}T_{L}^{b}T_{L}^{c}+\bar{\psi}%
_{R}^{\prime}\not A\psi_{R}^{\prime}C^{abc}T_{R}^{b}T_{R}^{c}\right)
\label{ct-no3Ab}%
\end{equation}%
\noindent
Summary

The sum of $\left(  \ref{ct-no3Aa}\right)  $ and $\left(  \ref{ct-no3Ab}%
\right)  $ is the total counter term due to the two diagrams in Figure
$\ref{bm-no3A}$:
\begin{equation}
\frac{1}{\left(  4\pi\right)  ^{2}}\frac{1}{6}g^{3}\left(  7+5\alpha\right)
\left(  \bar{\psi}_{L}\not A  \psi_{L}C_{L}T_{L}^{a}+\bar{\psi}_{R}^{\prime
}\not A  \psi_{R}^{\prime}C_{R}T_{R}^{a}\right)  \label{ct-no3A}%
\end{equation}

\begin{figure}
[tbh]
\begin{center}
\includegraphics[
height=0.7437in,
width=1.516in
]%
{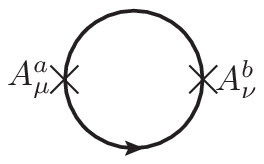}%
\caption{Diagrams for 2-point 1PI}%
\label{bm-nc2}%
\end{center}
\end{figure}

\subsubsection{Figure $\ref{bm-nc2}$: One-Fermion-Loop $2$-Point 1PI}%

Ignoring the non-Abelian group factor, the diagram in Figure $\ref{bm-nc2}$ is
the same as diagram $\left(  c\right)  $ in Figure $\ref{bm-c2}$. For the
$\psi$ fermion field, the group factor is $tr\left(  T_{L}^{a}T_{L}%
^{b}\right)  =T_{L}\delta^{ab}$. Consequently,%

\begin{equation}
\lim_{n\rightarrow4}\left(  \Gamma_{\psi,\ref{bm-nc2}}^{R5}-\Gamma
_{\psi,\ref{bm-nc2}}^{BM}\right)  =\frac{1}{\left(  4\pi\right)  ^{2}}%
ig^{\mu\nu}\frac{1}{3}g^{2}p^{2}T_{L}\delta^{ab}\nonumber
\end{equation}
where $p$ is the external momentum. The group factor for $\psi^{\prime}$ is
$tr\left(  T_{R}^{a}T_{R}^{b}\right)  =T_{R}\delta^{ab}$ which yields the
difference%
\begin{equation}
\lim_{n\rightarrow4}\left(  \Gamma_{\psi^{\prime},\ref{bm-nc2}}^{R5}%
-\Gamma_{\psi^{\prime},\ref{bm-nc2}}^{BM}\right)  =\frac{1}{\left(
4\pi\right)  ^{2}}ig^{\mu\nu}\frac{1}{3}g^{2}p^{2}T_{R}\delta^{ab}\nonumber
\end{equation}
The above two amplitudes can be accounted for by the counter term%
\begin{equation}
-\frac{1}{\left(  4\pi\right)  ^{2}}\frac{1}{6}g^{2}\left(  T_{L}%
+T_{R}\right)  A_{\mu}^{a}\square A^{a,\mu} \label{ct-nc2}%
\end{equation}

\begin{figure}
[tbh]
\begin{center}
\includegraphics[
height=1.1882in,
width=2.7657in
]%
{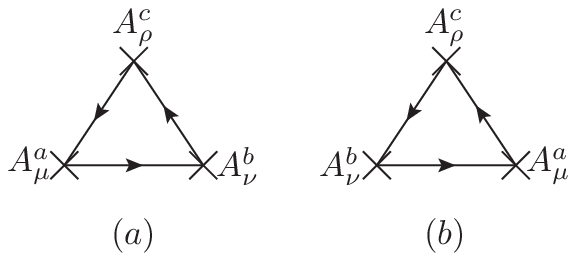}%
\caption{Diagrams for 3-point 1PI}%
\label{bm-nc3}%
\end{center}
\end{figure}

\subsubsection{Figure $\ref{bm-nc3}$: One-Fermion-Loop $3$-Point 1PI}%

We are only interested in terms that have an even count of $\gamma_{5}$.
Assume the incoming momenta entering $A_{\mu}^{a},A_{\nu}^{b},A_{\rho}^{c}$
are $k_{1},k_{2},k_{3}$.%

\noindent
Diagram $\left(  a\right)  $%
\begin{align*}
\Gamma_{\psi,\ref{bm-nc3}\left(  a\right)  }^{BM}  &  =-g^{3}\lim
_{m\rightarrow0}tr\int\left(  \frac{1}{\not \ell -m}R\gamma^{\mu}L\frac
{1}{\not \ell -\not k  _{1}-m}R\gamma^{\nu}L\frac{1}{\not \ell +\not k
_{3}-m}R\gamma^{\rho}L\right)  _{\gamma_{5}\text{-even}}\\
&  \times tr\left(  T^{a}T^{b}T^{c}\right)
\end{align*}%
\begin{align}
&  \lim_{n\rightarrow4}\left(  \Gamma_{\psi,\ref{bm-nc3}\left(  a\right)
}^{R5}-\Gamma_{\psi,\ref{bm-nc3}\left(  a\right)  }^{BM}\right) \label{fn3a1}%
\\
&  =-\frac{g^{3}}{2}\lim_{m\rightarrow0}\int\frac{tr\left(  \not \ell
\gamma^{\mu}\left(  \not \ell -\not k  _{1}\right)  \gamma^{\nu}\left(
\not \ell +\not k  _{3}\right)  \gamma^{\rho}-\underline{\not \ell }%
\gamma^{\mu}\left(  \underline{\not \ell }-\not k  _{1}\right)  \gamma^{\nu
}\left(  \underline{\not \ell }+\not k  _{3}\right)  \gamma^{\rho}\right)
}{\left(  \ell^{2}-m^{2}\right)  \left(  \left(  \ell-k_{1}\right)  ^{2}%
-m^{2}\right)  \left(  \left(  \ell+k_{3}\right)  ^{2}-m^{2}\right)
}\nonumber\\
&  \times tr\left(  T^{a}T^{b}T^{c}\right) \nonumber\\
&  =-\frac{1}{\left(  4\pi\right)  ^{2}}\frac{2i}{3}g^{3}\left(  \left(
k_{2}-k_{3}\right)  ^{\mu}g^{\nu\rho}+\left(  k_{3}-k_{1}\right)  ^{\nu}%
g^{\mu\rho}+\left(  k_{1}-k_{2}\right)  ^{\rho}g^{\mu\nu}\right)  tr\left(
T_{L}^{a}T_{L}^{b}T_{L}^{c}\right) \nonumber
\end{align}
Similarly,%
\begin{align}
&  \lim_{n\rightarrow4}\left(  \Gamma_{\psi^{\prime},\ref{bm-nc3}\left(
a\right)  }^{R5}-\Gamma_{\psi^{\prime},\ref{bm-nc3}\left(  a\right)  }%
^{BM}\right) \label{fn3a2}\\
&  =-\frac{1}{\left(  4\pi\right)  ^{2}}\frac{2i}{3}g^{3}\left(  \left(
k_{2}-k_{3}\right)  ^{\mu}g^{\nu\rho}+\left(  k_{3}-k_{1}\right)  ^{\nu}%
g^{\mu\rho}+\left(  k_{1}-k_{2}\right)  ^{\rho}g^{\mu\nu}\right)  tr\left(
T_{R}^{a}T_{R}^{b}T_{R}^{c}\right) \nonumber
\end{align}%
\noindent
Diagram $\left(  b\right)  $

Diagram $\left(  b\right)  $ can be obtained from diagram $\left(  a\right)  $
by the interchange $\left(  a,\mu,k_{1}\right)  \longleftrightarrow\left(
b,\nu,k_{2}\right)  $%
\begin{align}
&  \lim_{n\rightarrow4}\left(  \Gamma_{\psi,\ref{bm-nc3}\left(  b\right)
}^{R5}-\Gamma_{\psi,\ref{bm-nc3}\left(  b\right)  }^{BM}\right)
\label{fn3b1}\\
&  =\frac{1}{\left(  4\pi\right)  ^{2}}\frac{2i}{3}g^{3}\left(  \left(
k_{2}-k_{3}\right)  ^{\mu}g^{\nu\rho}+\left(  k_{3}-k_{1}\right)  ^{\nu}%
g^{\mu\rho}+\left(  k_{1}-k_{2}\right)  ^{\rho}g^{\mu\nu}\right)  tr\left(
T_{L}^{b}T_{L}^{a}T_{L}^{c}\right)  \nonumber
\end{align}%
\begin{align}
&  \lim_{n\rightarrow4}\left(  \Gamma_{\psi^{\prime},\ref{bm-nc3}\left(
b\right)  }^{R5}-\Gamma_{\psi^{\prime},\ref{bm-nc3}\left(  b\right)  }%
^{BM}\right)  \label{fn3b2}\\
&  =\frac{1}{\left(  4\pi\right)  ^{2}}\frac{2i}{3}g^{3}\left(  \left(
k_{2}-k_{3}\right)  ^{\mu}g^{\nu\rho}+\left(  k_{3}-k_{1}\right)  ^{\nu}%
g^{\mu\rho}+\left(  k_{1}-k_{2}\right)  ^{\rho}g^{\mu\nu}\right)  tr\left(
T_{R}^{b}T_{R}^{a}T_{R}^{c}\right)  \nonumber
\end{align}
Utilizing $tr\left(  \left[  T_{L}^{a},T_{L}^{b}\right]  T_{L}^{c}\right)
=iT_{L}C^{abc}$ and $tr\left(  \left[  T_{R}^{a},T_{R}^{b}\right]  T_{R}%
^{c}\right)  =iT_{R}C^{abc}$, the summation from $\left(  \ref{fn3a1}\right)
$ to $\left(  \ref{fn3b2}\right)  $ can be written as
\[
\frac{1}{\left(  4\pi\right)  ^{2}}\frac{2}{3}g^{3}\left(  \left(  k_{2}%
-k_{3}\right)  ^{\mu}g^{\nu\rho}+\left(  k_{3}-k_{1}\right)  ^{\nu}g^{\mu\rho
}+\left(  k_{1}-k_{2}\right)  ^{\rho}g^{\mu\nu}\right)  C^{abc}\left(
T_{L}+T_{R}\right)
\]
which leads to the counter term%
\begin{equation}
-\frac{1}{\left(  4\pi\right)  ^{2}}\frac{2}{3}g^{3}\left(  T_{L}%
+T_{R}\right)  C^{abc}\left(  \partial^{\mu}A_{\nu}^{a}\right)  A_{\mu}%
^{b}A^{c,\nu}\label{ct-nc3}%
\end{equation}

\begin{figure}
[tbh]
\begin{center}
\includegraphics[
height=2.0228in,
width=3.5993in
]%
{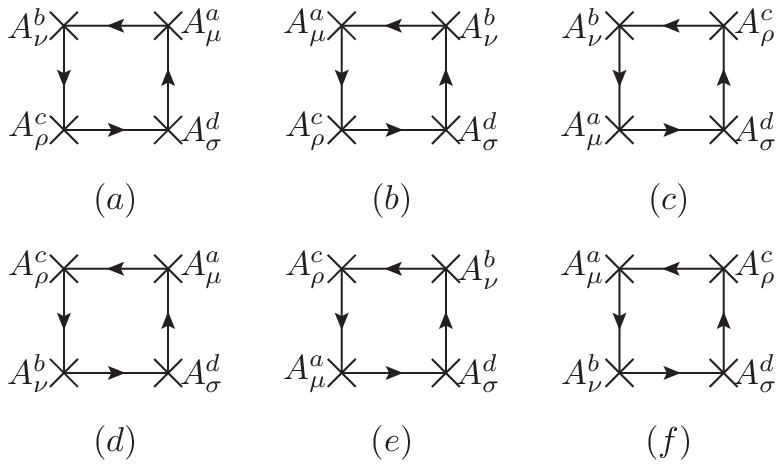}%
\caption{Diagrams for 4-point 1PI}%
\label{bm-nc4}%
\end{center}
\end{figure}

\subsubsection{Figure $\ref{bm-nc4}$: One-Fermion-Loop $4$-Point 1PI}%

\noindent
Diagram $\left(  a\right)  $

If $\psi$ is the fermion field, the amplitude for this diagram is equal to the
group factor $tr\left(  T_{L}^{a}T_{L}^{b}T_{L}^{c}T_{L}^{d}\right)
=T_{L}^{abcd}$ times the amplitude for diagram $\left(  i\right)  $ in Figure
$\ref{bm-c4}$. According to $\left(  \ref{fc4i}\right)  $, we then have
\begin{align}
&  \lim_{n\rightarrow4}\left(  \Gamma_{\psi,\ref{bm-nc4}\left(  a\right)
}^{R5}-\Gamma_{\psi,\ref{bm-nc4}\left(  a\right)  }^{BM}\right) \label{fnc4a1}%
\\
&  =\frac{1}{\left(  4\pi\right)  ^{2}}ig^{4}T_{L}^{abcd}\left(  g^{\mu\nu
}g^{\rho\sigma}-\frac{5}{3}g^{\mu\rho}g^{\nu\sigma}+g^{\mu\sigma}g^{\nu\rho
}\right) \nonumber
\end{align}
Likewise, if $\psi^{\prime}$ is the fermion field, $tr\left(  T_{R}^{a}%
T_{R}^{b}T_{R}^{c}T_{R}^{d}\right)  =T_{R}^{abcd}$ is the group factor and
\begin{align}
&  \lim_{n\rightarrow4}\left(  \Gamma_{\psi^{\prime},\ref{bm-nc4}\left(
a\right)  }^{R5}-\Gamma_{\psi^{\prime},\ref{bm-nc4}\left(  a\right)  }%
^{BM}\right) \label{fnc4a2}\\
&  =\frac{1}{\left(  4\pi\right)  ^{2}}ig^{4}T_{R}^{abcd}\left(  g^{\mu\nu
}g^{\rho\sigma}-\frac{5}{3}g^{\mu\rho}g^{\nu\sigma}+g^{\mu\sigma}g^{\nu\rho
}\right) \nonumber
\end{align}
The contribution to the difference from both $\psi$ and $\psi^{\prime}$
fermion loops is the sum of $\left(  \ref{fnc4a1}\right)  $ and $\left(
\ref{fnc4a2}\right)  $, which is equal to%
\begin{align}
&  \lim_{n\rightarrow4}\left(  \Gamma_{\ref{bm-nc4}\left(  a\right)  }%
^{R5}-\Gamma_{\ref{bm-nc4}\left(  a\right)  }^{BM}\right) \label{fnc4a}\\
&  =\frac{1}{\left(  4\pi\right)  ^{2}}ig^{4}\left(  T_{L}^{abcd}+T_{R}%
^{abcd}\right)  \left(  g^{\mu\nu}g^{\rho\sigma}-\frac{5}{3}g^{\mu\rho}%
g^{\nu\sigma}+g^{\mu\sigma}g^{\nu\rho}\right) \nonumber
\end{align}%
\noindent
Diagram $\left(  b\right)  $

The interchange $\left(  a,\mu\right)  \longleftrightarrow\left(
b,\nu\right)  $ on $\left(  \ref{fnc4a}\right)  $ yields
\begin{align}
&  \lim_{n\rightarrow4}\left(  \Gamma_{\ref{bm-nc4}\left(  b\right)  }%
^{R5}-\Gamma_{\ref{bm-nc4}\left(  b\right)  }^{BM}\right) \label{fnc4b}\\
&  =\frac{1}{\left(  4\pi\right)  ^{2}}ig^{4}\left(  T_{L}^{bacd}+T_{R}%
^{bacd}\right)  \left(  g^{\mu\nu}g^{\rho\sigma}+g^{\mu\rho}g^{\nu\sigma
}-\frac{5}{3}g^{\mu\sigma}g^{\nu\rho}\right) \nonumber
\end{align}%
\noindent
Diagram $\left(  c\right)  $

The interchange $\left(  a,\mu\right)  \longleftrightarrow\left(
c,\rho\right)  $ on $\left(  \ref{fnc4a}\right)  $ yields
\begin{align}
&  \lim_{n\rightarrow4}\left(  \Gamma_{\ref{bm-nc4}\left(  c\right)  }%
^{R5}-\Gamma_{\ref{bm-nc4}\left(  c\right)  }^{BM}\right) \label{fnc4c}\\
&  =\frac{1}{\left(  4\pi\right)  ^{2}}ig^{4}\left(  T_{L}^{cbad}+T_{R}%
^{cbad}\right)  \left(  g^{\mu\nu}g^{\rho\sigma}-\frac{5}{3}g^{\mu\rho}%
g^{\nu\sigma}+g^{\mu\sigma}g^{\nu\rho}\right) \nonumber
\end{align}%
\noindent
Diagram $\left(  d\right)  $

The interchange $\left(  b,\nu\right)  \longleftrightarrow\left(
c,\rho\right)  $ on $\left(  \ref{fnc4a}\right)  $ yields%
\begin{align}
&  \lim_{n\rightarrow4}\left(  \Gamma_{\ref{bm-nc4}\left(  d\right)  }%
^{R5}-\Gamma_{\ref{bm-nc4}\left(  d\right)  }^{BM}\right) \label{fnc4d}\\
&  =\frac{1}{\left(  4\pi\right)  ^{2}}ig^{4}\left(  T_{L}^{acbd}+T_{R}%
^{acbd}\right)  \left(  -\frac{5}{3}g^{\mu\nu}g^{\rho\sigma}+g^{\mu\rho}%
g^{\nu\sigma}+g^{\mu\sigma}g^{\nu\rho}\right) \nonumber
\end{align}%
\noindent
Diagram $\left(  e\right)  $

The interchange $\left(  b,\nu\right)  \longleftrightarrow\left(
c,\rho\right)  $ on $\left(  \ref{fnc4c}\right)  $ yields%
\begin{align}
&  \lim_{n\rightarrow4}\left(  \Gamma_{\ref{bm-nc4}\left(  e\right)  }%
^{R5}-\Gamma_{\ref{bm-nc4}\left(  e\right)  }^{BM}\right) \label{fnc4e}\\
&  =\frac{1}{\left(  4\pi\right)  ^{2}}ig^{4}\left(  T_{L}^{bcad}+T_{R}%
^{bcad}\right)  \left(  -\frac{5}{3}g^{\mu\nu}g^{\rho\sigma}+g^{\mu\rho}%
g^{\nu\sigma}+g^{\mu\sigma}g^{\nu\rho}\right) \nonumber
\end{align}%
\noindent
Diagram $\left(  f\right)  $

The interchange $\left(  a,\mu\right)  \longleftrightarrow\left(
c,\rho\right)  $ on $\left(  \ref{fnc4d}\right)  $ yields%
\begin{align}
&  \lim_{n\rightarrow4}\left(  \Gamma_{\ref{bm-nc4}\left(  f\right)  }%
^{R5}-\Gamma_{\ref{bm-nc4}\left(  f\right)  }^{BM}\right) \label{fnc4f}\\
&  =\frac{1}{\left(  4\pi\right)  ^{2}}ig^{4}\left(  T_{L}^{cabd}+T_{R}%
^{cabd}\right)  \left(  g^{\mu\nu}g^{\rho\sigma}+g^{\mu\rho}g^{\nu\sigma
}-\frac{5}{3}g^{\mu\sigma}g^{\nu\rho}\right) \nonumber
\end{align}
The summation from $\left(  \ref{fnc4a}\right)  $ to $\left(  \ref{fnc4f}%
\right)  $ gives
\begin{equation}
\frac{1}{\left(  4\pi\right)  ^{2}}ig^{4}\left(
\begin{array}
[c]{c}%
g^{\mu\nu}g^{\rho\sigma}\left(  \left(  T_{L+R}^{abcd}+T_{L+R}^{bacd}%
+T_{L+R}^{cbad}+T_{L+R}^{cabd}\right)  -\frac{5}{3}\left(  T_{L+R}%
^{acbd}+T_{L+R}^{bcad}\right)  \right) \\
+g^{\mu\rho}g^{\nu\sigma}\left(  \left(  T_{L+R}^{bacd}+T_{L+R}^{cabd}%
+T_{L+R}^{acbd}+T_{L+R}^{bcad}\right)  -\frac{5}{3}\left(  T_{L+R}%
^{abcd}+T_{L+R}^{cbad}\right)  \right) \\
+g^{\mu\sigma}g^{\nu\rho}\left(  \left(  T_{L+R}^{abcd}+T_{L+R}^{cbad}%
+T_{L+R}^{acbd}+T_{L+R}^{bcad}\right)  -\frac{5}{3}\left(  T_{L+R}%
^{bacd}+T_{L+R}^{cabd}\right)  \right)
\end{array}
\right)  \label{fnc4}%
\end{equation}
where $T_{L+R}^{abcd}=T_{L}^{abcd}+T_{R}^{abcd}$. The above amplitude can be
accounted for by the counter term%
\begin{align}
&  \frac{1}{\left(  4\pi\right)  ^{2}}g^{4}\left(  \frac{1}{2}T_{L+R}%
^{abcd}A^{a,\mu}A_{\mu}^{b}A^{c,\nu}A_{\nu}^{d}-\frac{5}{12}T_{L+R}%
^{abcd}A^{a,\mu}A^{b,\nu}A_{\mu}^{c}A_{\nu}^{d}\right) \label{ct-nc4}\\
&  =\frac{1}{\left(  4\pi\right)  ^{2}}g^{4}\left(  \frac{1}{12}T_{L+R}%
^{abcd}A^{a,\mu}A_{\mu}^{b}A^{c,\nu}A_{\nu}^{d}+\frac{5}{24}\left(
T_{L}+T_{R}\right)  C^{eab}C^{ecd}A_{\mu}^{a}A_{\nu}^{b}A^{c,\mu}A^{d,\nu
}\right) \nonumber
\end{align}

\subsubsection{One-Loop Counter Terms for the Non-Abelian Theory}

The results for the finite counter terms stemming from the difference of
amplitudes between the rightmost scheme and the BM scheme calculated for the
diagrams in Figure $\ref{bm-no2}$-$\ref{bm-nc4}$ in the chiral non-Abelian
gauge theory are summarized in Table \ref{ct-nab}.%

\begin{table}[htbp] \centering
\begin{tabular}
[c]{|l|l|l|}\hline
Figure & where & $\left(  4\pi\right)  ^{2}\times$ Counter Term\\\hline
\ref{bm-no2} & $\left(  \ref{ct-no2}\right)  $ & $-\frac{1}{3}g^{2}\left(
1+2\alpha\right)  \left(  \bar{\psi}_{L}i\not \partial \psi_{L}C_{L}+\bar
{\psi}_{R}^{\prime}i\not \partial \psi_{R}^{\prime}C_{R}\right)  $\\
\ref{bm-no3A} & $\left(  \ref{ct-no3A}\right)  $ & $\frac{1}{6}g^{3}\left(
7+5\alpha\right)  \left(  \bar{\psi}_{L}\not A  \psi_{L}C_{L}T_{L}^{a}%
+\bar{\psi}_{R}^{\prime}\not A  \psi_{R}C_{R}^{\prime}T_{R}^{a}\right)  $\\
\ref{bm-nc2} & $\left(  \ref{ct-nc2}\right)  $ & $-\frac{1}{6}g^{2}\left(
T_{L}+T_{R}\right)  A_{\mu}^{a}\square A^{a,\mu}$\\
\ref{bm-nc3} & $\left(  \ref{ct-nc3}\right)  $ & $-\frac{2}{3}g^{3}\left(
T_{L}+T_{R}\right)  C^{abc}\left(  \partial^{\mu}A_{\nu}^{a}\right)  A_{\mu
}^{b}A^{c,\nu}$\\
\ref{bm-nc4} & $\left(  \ref{ct-nc4}\right)  $ & $\left(
\begin{array}
[c]{c}%
\frac{1}{12}g^{4}\left(  T_{L}^{abcd}+T_{R}^{abcd}\right)  A^{a,\mu}A_{\mu
}^{b}A^{c,\nu}A_{\nu}^{d}\\
+\frac{5}{24}g^{4}\left(  T_{L}+T_{R}\right)  C^{eab}C^{ecd}A_{\mu}^{a}A_{\nu
}^{b}A^{c,\mu}A^{d,\nu}%
\end{array}
\right)  $\\\hline
\end{tabular}
\caption{Counter terms due to diagrams in Figure \ref{bm-no2}-\ref{bm-nc4}\label{ct-nab}}\label{tb8}%
\end{table}%


\begin{thebibliography}{99}                                                                                               %


\bibitem {HV}G. 't Hooft and M. Veltman, Nucl. Phys. \textbf{B44}, 189 (1972).

\bibitem {BM}P. Breitenlohner and D. Maison, Commun. math. Phys. \textbf{52},
11 (1977).

\bibitem {WTI}J. C. Ward, Phys. Rev. \textbf{78}, 182 (1950); Y. Takahashi,
Nuovo Cimento, \textbf{6}, 371 (1957).

\bibitem {BRST}C. Becchi, A. Rouet and R. Stora, Phys. Lett. \textbf{B52}, 344
(1974); Comm. Math. Phys. \textbf{42}, 127 (1975); Ann. of Phys. \textbf{98},
287 (1976). I.V. Tyutin, Lebedev Institute preprint 39 (1975).

\bibitem {GB2}Guy Bonneau, Nucl. Phys. \textbf{B177}, 523 (1981).

\bibitem {FER}R. Ferrari, A. Le Yaouanc, L. Oliver, and J. C. Raynal, Phys.
Rev. \textbf{D52}, 3036 (1995).

\bibitem {FG}R. Ferrari and P. A. Grassi, Phys. Rev. \textbf{D}60, 065010 (1999).

\bibitem {MS}C. P. Martin and D. Sanchez-Ruiz, Nucl. Phys. \textbf{B572}, 387 (2000).

\bibitem {SANR}D. Sanchez-Ruiz, Phys. Rev. \textbf{D68}, 025009 (2003).

\bibitem {ECSM}E. C. Tsai, Gauge Invariant Treatment of $\gamma_{5}$ in the
Scheme of 't Hooft and Veltman, arXiv:0905.1550v3 [hep-th].
\end{thebibliography}
\end{document}